\documentclass{article}

\usepackage{arxiv}
\usepackage[utf8]{inputenc}
\usepackage[T1]{fontenc}
\usepackage{amsthm}

\newif\ifPolyRefUsesBm
\PolyRefUsesBmtrue
\newcommand{\PaperTitle}{Tail variation and separation in reference-ratio categorical regression}
\newcommand{\PaperShortTitle}{Tail variation and separation}
\newcommand{\PaperAuthors}{Lukas Sablica; Kurt Hornik; Thomas Rusch}
\newcommand{\PaperShortAuthors}{Sablica et al.}
\newcommand{\PaperKeywords}{Categorical regression; Concavity; Maximum Likelihood; Reference-Ratio Model; Regular Variation; Separation.}

\newcommand{\GenerativeAIDeclaration}{%
\section*{Declaration of the use of generative AI and AI-assisted technologies}

During the preparation of this work the authors used OpenAI GPT 5.5 to assist with spell-checking, final language review and polishing. After using these tools the authors reviewed and edited the content as necessary and take full responsibility for the content of the publication.
}

\newcommand{\StatsInstitution}{Institute for Statistics and Mathematics}
\newcommand{\MethodsInstitution}{Competence Center for Empirical Research Methods}
\newcommand{\PaperUniversity}{WU Vienna University of Economics and Business}
\newcommand{\PaperAddress}{Welthandelsplatz 1, 1020 Vienna}
\newcommand{\PaperCountry}{Austria}

\newcommand{\LukasEmail}{Lukas.Sablica@wu.ac.at}
\newcommand{\LukasORCID}{0000-0001-9166-4563}
\newcommand{\KurtEmail}{Kurt.Hornik@wu.ac.at}
\newcommand{\KurtORCID}{0000-0003-4198-9911}
\newcommand{\ThomasEmail}{Thomas.Rusch@wu.ac.at}
\newcommand{\ThomasORCID}{0000-0002-7773-2096}

\usepackage{amsmath}
\usepackage{amssymb}
\ifPolyRefUsesBm
\usepackage{bm}
\fi
\usepackage{booktabs}
\usepackage{graphicx}
\usepackage{newtxtext}
\usepackage[subscriptcorrection]{newtxmath}

\def\T{{\mathrm{\scriptscriptstyle T}}}
\newcommand{\pkg}[1]{{\normalfont\fontseries{b}\selectfont #1}}
\newcommand{\R}{\mathbb{R}}

\newcommand{\E}{\mathbb{E}}

\newcommand{\calS}{\mathcal{S}}
\newcommand{\calX}{\mathcal{X}}
\newcommand{\lodds}{\lambda}
\newcommand{\loglik}{\ell}

\providecommand{\qedhere}{}

\newif\ifPolyRefAltText
\newcommand{\alttext}[1]{\ifPolyRefAltText\par\vspace{2pt}\noindent Alt text: #1\fi}

\usepackage[round]{natbib}
\usepackage[colorlinks=true,linkcolor=blue,citecolor=blue,urlcolor=blue,hypertexnames=false]{hyperref}

\theoremstyle{plain}
\newtheorem{theorem}{Theorem}
\newtheorem{lemma}{Lemma}
\newtheorem{proposition}{Proposition}
\newtheorem{corollary}{Corollary}
\newtheorem{definition}{Definition}
\newdimen\figwidth

\NewCommandCopy\amsthmproof\proof
\NewCommandCopy\endamsthmproof\endproof
\renewenvironment{proof}[1][]{%
  \if\relax\detokenize{#1}\relax
    \amsthmproof
  \else
    \amsthmproof[\proofname\ #1]%
  \fi}{\endamsthmproof}

\newenvironment{restated}[1]{\par\addvspace{\medskipamount}\noindent{\normalfont\bfseries #1.}\ \itshape\ignorespaces}{\par\addvspace{\medskipamount}}

\newcommand{\ReplicationLocation}{is provided as the ancillary file \texttt{polyref.R} with the arXiv version of this paper}
\renewcommand{\headeright}{\PaperShortAuthors}
\renewcommand{\undertitle}{A Preprint}
\title{\PaperTitle}
\author{
  Lukas Sablica \\
  \StatsInstitution\\
  \PaperUniversity\\
  \PaperAddress, \PaperCountry\\
  ORCID: \LukasORCID\\
  \href{mailto:\LukasEmail}{\nolinkurl{\LukasEmail}}
  \And
  Kurt Hornik\\
  \StatsInstitution\\
  \PaperUniversity\\
  \PaperAddress, \PaperCountry\\
  ORCID: \KurtORCID\\
  \href{mailto:\KurtEmail}{\nolinkurl{\KurtEmail}}
  \And
  Thomas Rusch\\
  \MethodsInstitution\\
  \PaperUniversity\\
  \PaperAddress, \PaperCountry\\
  ORCID: \ThomasORCID\\
  \href{mailto:\ThomasEmail}{\nolinkurl{\ThomasEmail}}
}
\date{}

\hypersetup{
  pdftitle={\PaperTitle},
  pdfauthor={\PaperAuthors}
}

\begin{document}

\maketitle
\fancyhead[C]{\PaperShortTitle}

\begin{abstract}
In categorical regression, separation is often taken to rule out a finite maximum likelihood estimator and overlap to guarantee one, regardless of the link. We show that this reading fails for a very general class of reference-ratio models, which extend a binary link to nominal responses through the odds transform of its distribution function. These models have a score-transformed softmax representation. Existence depends jointly on the separation geometry and the tail variation of the odds transform. In fixed-reference models with the rapidly varying logit, probit and complementary log-log links, complete separation precludes a finite maximizer and overlap ensures one. Among these links, only logit guarantees nonexistence under quasi-complete separation. Regularly varying transforms, including the odds transforms of the Cauchit and Student t links, introduce an intermediate plateau regime in which pairwise separation geometry is nondecisive in both directions. In this regime, completely separated data may admit a finite maximizer outside the separating cone, while data exhibiting overlap may fail to have any finite maximizer. Furthermore, we prove that the multinomial logit is the unique globally concave member of the full normalized-coordinate reference-ratio family. This global concavity implies exponential tail sensitivity, and heavy-tailed links gain robustness to separation only by departing from the globally concave exponential-family class.

\end{abstract}


\section{Introduction}

Categorical regression models translate linear predictors into probabilities on a finite response set, with standard treatments covering nominal, ordinal and discrete-choice data \citep{polytomous:Fahrmeir+Tutz:1994,Agresti:2002,Tutz:2012}. Classical families include cumulative links \citep{McCullagh}, adjacent-category links \citep{polytomous:Simon:1974,polytomous:Andrich:1978,polytomous:Agresti:2010}, continuation-ratio or sequential links \citep{polytomous:Thompson:1977,polytomous:Tutz:1991}, stereotype models \citep{polytomous:Anderson:1984}, and random-utility choice models \citep{mcfadden1974,Train:2009}. For nominal responses, the multinomial logit is the default because it is simple, globally concave, and has a random-utility interpretation. Yet separation can preclude a finite maximum likelihood estimator. In binary logistic regression, separation can also leave exact conditional confidence intervals unbounded \citep{lewis-battey-2024}.

The separation problem has a long history in generalized linear and categorical response models. Early work identified the geometry of separation and overlap in binary and multinomial logistic models \citep{polytomous:Wedderburn:1976,polytomous:Silvapulle:1981,polytomous:Albert+Anderson:1984,Santner+Duffy:1986,separation:Lesaffre-Albert:1989}. Related existence results were then developed for quantal, ordinal, probit and polytomous models \citep{polytomous:Kaufmann:1988,polytomous:Tse:1986,probit:Lesaffre-Kaufmann:1992,polytomous:Chen+Shao:2000,polytomous:Speckman+Lee+Sun:2009}. Penalized and bias-reducing estimators can remain finite under separation \citep{firth,Heinze+Schemper:2002,kosmidis-lunardon-2024}, and software diagnostics exploit the underlying convex geometry \citep{detectseparation,divorce}. Recent work by \citet{sablica-hornik-rusch-crm} gives a very general geometric treatment of existence for the usual random-utility and category-ratio constructions. The reference-ratio models of \citet{peyhardi+trottier+guedon:2015}, implemented for example in \pkg{GLMcat} \citep{glmcat}, are different. Except in the logistic case, their probability ratios generally depend on the two linear predictors separately. The usual separation geometry is therefore only part of the story.

In reference-ratio categorical regression, a continuous distribution function $F$ used as a binary-response link is extended to $J$ nominal categories by normalizing its centered odds transform $\tau$. Write $\tau_{\rm raw}=F/(1-F)$ for the raw odds. The likelihood at infinity is governed by ratios of centered odds evaluated at the category predictors. Hence the consequences of separation depend on the tail variation of $\tau$. Rapidly varying odds, such as those produced by logit, probit and complementary log-log links, make strict slope advantages overwhelming. For regularly varying odds, such as those produced by Cauchit and Student $t$ links, a positive asymptotic contribution can remain even under a strict slope advantage when both predictors diverge in the same tail. This produces an intermediate plateau regime in which pairwise separation geometry is nondecisive in both directions. Separated data may admit a finite maximizer, while data with overlap may yield an unattained boundary supremum. Changing the regular-variation index, the exponent governing power-law growth of the right-tail odds or decay of the left-tail odds, can reverse existence on the same data.

The result is a sharp tradeoff. Classical work on binary-response and cumulative-link models relates likelihood concavity to link choice \citep{Pratt:1981,McCullagh}. In the full normalized-coordinate formulation, the multinomial logit (standard softmax) is the unique reference-ratio model whose log-likelihood is globally concave for every design. Equivalently, global concavity on that unrestricted coordinate space forces affine log odds. Affine log odds imply exponential tail sensitivity, so heavy-tailed links gain robustness to separated data at the cost of leaving the globally concave exponential-family class. This reflects a structural incompatibility between global concavity and heavy-tail suppression.

We consider the reference-ratio class of categorical response models \citep{glmcat,peyhardi+trottier+guedon:2015}. Take $F$ to be continuous and strictly increasing on $\R$ and let $s_0=\tau_{\rm raw}^{-1}(1)$. We work with the centered odds transform $\tau(s)=\tau_{\rm raw}(s+s_0)$, so $\tau(0)=1$. This is a horizontal recentering of the binary link and leaves right and left tail variation unchanged. It includes asymmetric links such as complementary log-log without requiring $F(0)=1/2$.

For categories $1,\ldots,J$, let the last category be the reference category and set
\[
  \gamma_{iJ}=0,\qquad \tau(\gamma_{iJ})=1.
\]
We call this specification, in which the reference predictor is identically zero, the fixed-reference formulation. For $j=1,\ldots,J-1$, write $\gamma_{ij}=z_{ij}^\T\beta$, where $z_{ij}\in\R^p$. The response probabilities are
\[
  h_{ij}(\beta)
  =
  \frac{\tau(\gamma_{ij})}{1+\sum_{k=1}^{J-1}\tau(\gamma_{ik})}
  \quad (j=1,\ldots,J-1),
  \qquad
  h_{iJ}(\beta)
  =
  \frac{1}{1+\sum_{k=1}^{J-1}\tau(\gamma_{ik})}.
\]
Equivalently, with $z_{iJ}=0$ this can be written as
\[
  h_{ij}(\beta)=\frac{\tau(z_{ij}^\T\beta)}
  {\sum_{k=1}^J\tau(z_{ik}^\T\beta)}
  \quad (j=1,\ldots,J),
  \qquad \tau(0)=1.
\]
The centered odds transform
\[
  \tau(s) = \frac{F(s+s_0)}{1-F(s+s_0)}
\]
is formed from a strictly increasing distribution function $F$ on $\mathbb{R}$, so $\tau(s)\to 0$ as $s\to -\infty$ and $\tau(s)\to\infty$ as $s\to\infty$.

For observed responses $y_1,\ldots,y_n$, we write the likelihood and log-likelihood as
\[
  L(\beta)=\prod_{i=1}^n h_{i,y_i}(\beta),
  \qquad
  \loglik(\beta)=\sum_{i=1}^n \log h_{i,y_i}(\beta).
\]
If $F=\mathrm{logit}^{-1}$ (the standard logistic distribution function), then $\tau(s)=e^s$ and we recover the multinomial logit probabilities
\[
  h_{ij}(\beta) = \frac{e^{z_{ij}^\T\beta}}{\sum_{k=1}^J e^{z_{ik}^\T\beta}}.
\]
Other choices of $F$ lead to different tail behaviors.

The reference-ratio probabilities have a softmax representation in which a common increasing transformation $\lodds=\log\tau$ is applied to each linear predictor before exponentiation and normalization. In this representation,
\[
  h_{ij}(\beta)
  =
  \frac{\exp\{\lodds(z_{ij}^\T\beta)\}}
  {\sum_{k=1}^J\exp\{\lodds(z_{ik}^\T\beta)\}}.
\]
The tail results thus apply to the corresponding softmax regression models with a common score transformation and a fixed reference. This also gives a Gumbel choice model with transformed systematic utilities. The logit link has $\lodds(s)=s$, while Cauchit and Student $t$ links replace linear utility growth by logarithmic growth in the tails. This representation gives reference-ratio links a simple nominal-response form, with closed-form normalized probabilities and tail-controlled sensitivity to separated observations.

A convenient reformulation of the model is obtained by dividing numerator and denominator by \(\tau(\gamma_{ij})\):
\[
  h_{ij}(\beta) = \frac{1}{1 + \sum_{k\ne j} \frac{\tau(z_{ik}^\T\beta)}{\tau(z_{ij}^\T\beta)}}.
\]
The tail behavior of the centered odds transform controls whether these probability ratios vanish, diverge or converge to positive constants along rays.

Proofs of all lemmas, and of Propositions~\ref{prop:rr-ray-completion} and~\ref{prop:fixedref}, are given in the Supplementary Material.

\section{Rapid variation}

\begin{definition}[Rapid variation at $+\infty$]
Let $\tau:\mathbb R\to(0,\infty)$ be the centered odds transform and write $R_{ab}(t)=\tau(bt)/\tau(at)$ for slopes $a,b$. We say that its right tail is rapidly varying if the restriction of $\tau$ to $[0,\infty)$ is ultimately increasing and for any $a>b\ge 0$,
\begin{equation}
R_{ab}(t) \xrightarrow[t\to\infty]{} 0.
\label{eq:RV-scale}
\end{equation}
In particular, rapid variation implies that $\tau(t)$ grows faster than any power of $t$,
\[
\lim_{t\to\infty} \frac{\tau(t)}{t^M} = \infty \quad \text{for every fixed } M>0.
\]
\end{definition}

A commonly used sufficient condition is the additive step growth
\begin{equation}
\exists\Delta>0, c>1, S \quad \text{such that}\quad \frac{\tau(s+\Delta)}{\tau(s)} \,\ge\, c\qquad \forall s\ge S.
\label{eq:RV-step}
\end{equation}
One always has \eqref{eq:RV-step} $\Rightarrow$ \eqref{eq:RV-scale}, but not conversely in general. For example,
$\tau(s)=\exp(\sqrt{s})$ satisfies \eqref{eq:RV-scale}, yet
$\tau(s+\Delta)/\tau(s)\to 1$ for any fixed $\Delta>0$, so \eqref{eq:RV-step} fails.

\begin{lemma}[Step $\Rightarrow$ scaling]\label{lem:step-scaling}
If $\tau$ satisfies \eqref{eq:RV-step} and is ultimately increasing, then $\tau$ is rapidly varying in the sense of \eqref{eq:RV-scale}.
\end{lemma}

\begin{definition}[Rapidly varying model]
A model in this family is rapidly varying if its odds transform $\tau$ is rapidly varying at $+\infty$, in the sense of \eqref{eq:RV-scale}, and its left-tail transform $\tilde{\tau}(s) = 1/\tau(-s)$ is also rapidly varying at $+\infty$.
\end{definition}

For standard links the classification follows from elementary tail expansions. Let $\bar F=1-F$ be the survival function. For the centered odds transform,
\[
\tau(s)=\frac{F(s+s_0)}{1-F(s+s_0)}
=\frac{1}{\bar F(s+s_0)}-1
\sim \frac{1}{\bar F(s+s_0)}
\qquad (s\to+\infty).
\]
Thus the odds tail is asymptotically the reciprocal of the survival probability. For logit, $\tau(s)=e^s$, so $\tau(s+\Delta)/\tau(s)=e^\Delta$ and \eqref{eq:RV-step} holds. For probit, Mills' ratio gives $\bar\Phi(s)\sim\phi(s)/s$, hence $\tau(s)\sim \sqrt{2\pi}\,s e^{s^2/2}$ and $\tau(s+\Delta)/\tau(s)\to\infty$ for every $\Delta>0$. For complementary log-log, before recentering, $\bar F(s)=e^{-e^s}$ and $\tau_{\rm raw}(s)\sim e^{e^s}$, so $\tau_{\rm raw}(s+\Delta)/\tau_{\rm raw}(s)\sim \exp\{(e^\Delta-1)e^s\}\to\infty$. Its left tail satisfies $\tau_{\rm raw}(s)\sim e^s$ as $s\to-\infty$, and therefore $1/\tau_{\rm raw}(-u)\sim e^u$ as $u\to\infty$. For symmetric links such as logit and probit, the left-tail transform is the same as the right-tail transform. Hence logit, probit and complementary log-log links are rapidly varying in the sense above. Cauchit and Student $t$ links are not, and are treated in Section~\ref{sec:regular}.

\begin{definition}[Estimable span and pairwise separation]\label{def:span-sep}
Here and below, $[J]=\{1,\ldots,J\}$ denotes the category index set.
Let the structure set be
\[
\calS = \{ s_{i,k} = z_{i,y_i}-z_{ik} \mid i=1,\dots,n,\ k\in[J]\setminus\{y_i\}\} \subset \mathbb R^p.
\]
The estimable span is
\[
\calX=\mathrm{span}\{z_{ij}\mid i=1,\ldots,n,\ j=1,\ldots,J\}.
\]
Every $\beta\in\R^p$ decomposes as $\beta=\beta_{\calX}+\beta_{\perp}$ with $\beta_{\calX}\in\calX$ and $\beta_\perp\perp\calX$. Since $z_{ij}^\T\beta_\perp=0$, the likelihood depends only on $\beta_{\calX}$. All separation conditions below are therefore defined for nonzero directions $v\in\calX$.

We say that $\calS$ exhibits complete separation if there exists $v\in\calX\setminus\{0\}$ with
\[
s^\T v > 0 \quad \text{for all } s\in\calS.
\]
We say $\calS$ has quasi-complete separation if there exists $v\in\calX\setminus\{0\}$ such that $s^\T v \ge 0$ for all $s\in\calS$ and at least one strict inequality holds. Otherwise $\calS$ has overlap.
\end{definition}

All maximization and limiting arguments are henceforth taken over $\calX$. This convention removes rank assumptions without changing the likelihood surface.

\subsection{Complete separation under rapid variation}

\begin{lemma}[Asymptotic ratios for rapidly varying models]\label{lem:rv-ratio-limit}
For a rapidly varying model, let $a, b \in \mathbb{R}$ satisfy $a > b$, and let $c,d\in \R$. Then
\[
\lim_{t\to\infty} \frac{\tau(b t + d)}{\tau(a t + c)} = 0.
\]
\end{lemma}

\begin{proposition}[Complete separation]\label{prop:complete-sep}
Consider a rapidly varying reference-ratio model. If $\calS$ exhibits complete separation in $\calX$, then no finite maximum likelihood estimator exists.
\end{proposition}
\begin{proof}
Let \(v\in\mathcal X\setminus\{0\}\) satisfy complete separation and fix any
finite \(\beta_0\). Set \(\beta(t)=\beta_0+tv\), let \(j=y_i\), and write \(a_{ik}=z_{ik}^{\T}v\), \(b_{ik}=z_{ik}^{\T}\beta_0\).

Complete separation gives \(a_{ij}>a_{ik}\) for every \(i\) and every
\(k\ne j\). Hence Lemma~\ref{lem:rv-ratio-limit} gives
\[
\frac{\tau(z_{ik}^{\T}\beta(t))}
     {\tau(z_{ij}^{\T}\beta(t))}
=
\frac{\tau(a_{ik}t+b_{ik})}{\tau(a_{ij}t+b_{ij})}
\longrightarrow 0,
\qquad k\ne j .
\]
Therefore \(h_{ij}(\beta(t))\to1\) for every \(i\), and consequently
\(L(\beta(t))\to1\). Since all denominator terms are strictly positive,
\(h_{i,y_i}(\beta)<1\) for every finite \(\beta\), and hence \(L(\beta)<1\)
for every finite \(\beta\). Thus the supremum is approached only along a
diverging ray and is not attained at any finite point. No finite maximum
likelihood estimator exists.
\end{proof}

\subsection{Nondecisive quasi-separation}

For rapidly varying reference-ratio models, quasi-complete pairwise separation need not preclude a finite maximum likelihood estimator. This distinguishes the general reference-ratio family from the random-utility and category-ratio models studied by \citet{sablica-hornik-rusch-crm}.

The problem lies with observations $i$ whose slopes are ``tied'' along the separating ray $v$. Quasi-complete separation, $s_{i,k}^\T v = (z_{ij}-z_{ik})^\T v \ge 0$, only guarantees $a_{ij} \ge a_{ik}$ for all $i, k \ne j$. It allows for the possibility of ties, $a_{ij} = a_{ik} = A$, for many observations, so long as at least one strict inequality $a_{i^*j^*} > a_{i^*k^*}$ holds.

The behavior of these tied ratios $R_{ik}(t) = \tau(At+b_{ik}) / \tau(At+b_{ij})$ is the source of the nondecisive separation phenomenon. Let $\lodds = \log \tau$. By the mean value theorem,
\[
\log R_{ik}(t) = \lodds(At+b_{ik}) - \lodds(At+b_{ij}) = \lodds'(\xi_t)(b_{ik}-b_{ij}),
\]
where $\xi_t$ is between $At+b_{ij}$ and $At+b_{ik}$.

For logit ($\tau(s)=e^s$), tied ratios remain constant at $e^{b_{ik}-b_{ij}}$, while the ratios for strictly dominated competitors decrease to zero. Thus the likelihood increases strictly along every quasi-separating ray, so no finite maximizer exists.

For probit and complementary log-log, $\lodds'(s)\to\infty$ as $s\to+\infty$. If $A>0$, the sign of the fixed intercept gap $\Delta_{ik}:=b_{ik}-b_{ij}$ determines the limit:
\[
\Delta_{ik} > 0 \implies R_{ik}(t) \to \infty, \qquad \Delta_{ik} < 0 \implies R_{ik}(t) \to 0.
\]
For complementary log-log, the centered transform satisfies $\tau(s)\sim\exp(c_0e^s)$ with $c_0=e^{s_0}=\log 2$. The tied ratio is asymptotic to $\exp\!\big\{c_0e^{At}(e^{b_{ik}}-e^{b_{ij}})\big\}$, giving the same limits.

Thus a quasi-separating ray $v$ makes ratios for strictly dominated competitors tend to zero, while tied ratios can increase or decrease. If a tied row $i$ has an unfavorable intercept gap at $\beta_0$, namely $b_{ik} > b_{ij}$ for some $k \ne j$ and $A>0$, then $R_{ik}(t)$ diverges and $h_{ij}(\beta(t)) \to 0$. A single tied row can therefore dominate the likelihood along that ray.

At a neutral start $\beta_0=0$, every tied ratio equals $1$ and each strictly dominated competitor's ratio tends to zero. If the ray has a tied row, the likelihood tends to a plateau $L_\infty(0,v)\in(0,1)$, where $L_\infty(\beta_0,v)=\lim_{t\to\infty}L(\beta_0+tv)$. A finite likelihood value may exceed this limit.

Section~S2.1 of the Supplementary Material gives a two-observation fixed-reference example with quasi-complete separation for which both the probit and complementary log-log likelihoods attain finite maxima, even after all divergent sequences are taken into account. Thus quasi-complete separation is decisive for logit, but not for probit or complementary log-log. The following one-vs-all condition suffices for every rapidly varying model.

\begin{proposition}[Quasi-one-vs-all separation]\label{prop:nn-sep}
Consider a rapidly varying reference-ratio model on $\calX$. Let $j=y_i$ be the observed category for observation $i$. Suppose there exists a vector $v \in\calX\setminus\{0\}$ such that for every observation $i$:
\[
z_{ij}^\T v \ge 0 \quad \text{and} \quad z_{ik}^\T v \le 0 \quad \text{for all } k \ne j.
\]
Then no finite maximum likelihood estimator exists.
\end{proposition}

\begin{proof}
Let $\beta_0$ be any finite parameter vector. We analyze the likelihood $L(\beta(t))$ along the ray $\beta(t) = \beta_0 + t v$ for $t \to \infty$. For observation $i$, with $j=y_i$, write $a_{ik}=z_{ik}^\T v$, $b_{ik}=z_{ik}^\T\beta_0$ and $R_{ik}(t)=\tau(a_{ik}t+b_{ik})/\tau(a_{ij}t+b_{ij})$.

For any observation $i$ where all slopes $a_{ik}$ are zero, its probability contribution $h_{ij}(\beta(t)) = \frac{\tau(b_{ij})}{\sum_k \tau(b_{ik})}$ is constant in $t$.

Because $v\in\calX\setminus\{0\}$, at least one slope $z_{ik}^\T v$ is nonzero. For any such strict row $i^*$, with $j^*=y_{i^*}$, the condition implies $a_{i^*j^*} \ge a_{i^*k}$ for all $k \ne j^*$, with at least one strict inequality. By Lemma \ref{lem:rv-ratio-limit}, any ratio $R_{i^*k^*}(t)$ where $a_{i^*j^*} > a_{i^*k^*}$ tends to zero. Other ratios with $a_{i^*j^*} = a_{i^*k^*}=0$ remain constant. Thus the limit of $1+\sum_{k\ne j^*}R_{i^*k}(t)$ is strictly smaller than its value at $t=0$. For all other rows, the limit of $1+\sum_{k\ne j}R_{ik}(t)$ is no larger than its value at $t=0$.

Consequently,
\[
\lim_{t\to\infty} L(\beta_0+tv) > L(\beta_0).
\]
For sufficiently large $t$, we have $L(\beta_0 + tv) > L(\beta_0)$, so no finite $\beta_0$ can be the maximizer. Thus, no finite MLE exists.
\end{proof}

Proposition \ref{prop:nn-sep} shows that the obstruction is not a tie itself, but a nonzero tie. Under rapid tails with increasing log odds derivative, a tie with $a_{ij}=a_{ik}=A\ne0$ can be decided by the starting intercepts. Under quasi-one-vs-all separation, all ties occur at zero slope, while at least one row changes because $v$ is nonzero in the estimable span.

\subsection{Existence under overlap}

\begin{proposition}[Overlap implies existence]\label{prop:rv-overlap}
Consider a rapidly varying fixed-reference model on \(\calX\). If $\calS$ exhibits overlap, then a finite maximum likelihood estimator exists.
\end{proposition}
\begin{proof}
It suffices to show that \(L(\beta)\to0\) as \(\|\beta\|\to\infty\) on
\(\calX\). Suppose otherwise. Then there are \(\varepsilon>0\) and
\(\beta_n\in\calX\), with \(\|\beta_n\|\to\infty\), such that
\(L(\beta_n)\ge\varepsilon\). Write \(\beta_n=t_n v_n\), where
\(t_n=\|\beta_n\|\to\infty\), \(v_n\in\calX\), and \(\|v_n\|=1\). Passing to a
subsequence, \(v_n\to v\in\calX\) with \(\|v\|=1\). Because \(z_{iJ}=0\), the structure vectors span \(\calX\), so \(v\) cannot be orthogonal to every element of \(\calS\).

By overlap, there exist an observation \(i^*\) and a competitor
\(k^*\ne y_{i^*}\) such that
\[
(z_{i^*,y_{i^*}}-z_{i^*,k^*})^\T v<0.
\]
Equivalently, with
\[
a=z_{i^*,k^*}^{\T}v,\qquad b=z_{i^*,y_{i^*}}^{\T}v,
\]
we have \(a>b\). Choose \(\eta>0\) such that \(a-\eta>b+\eta\). Since
\(v_n\to v\), for all sufficiently large \(n\),
\[
z_{i^*,k^*}^{\T}v_n\ge a-\eta,
\qquad
z_{i^*,y_{i^*}}^{\T}v_n\le b+\eta .
\]
Using monotonicity of \(\tau\) and Lemma~\ref{lem:rv-ratio-limit},
\[
\frac{\tau(z_{i^*,y_{i^*}}^{\T}\beta_n)}
     {\tau(z_{i^*,k^*}^{\T}\beta_n)}
\le
\frac{\tau((b+\eta)t_n)}{\tau((a-\eta)t_n)}
\longrightarrow 0 .
\]
Hence
\[
h_{i^*,y_{i^*}}(\beta_n)
\le
\frac{\tau(z_{i^*,y_{i^*}}^{\T}\beta_n)}
     {\tau(z_{i^*,k^*}^{\T}\beta_n)}
\longrightarrow 0.
\]
Since all likelihood contributions lie in \((0,1]\),
\(L(\beta_n)\le h_{i^*,y_{i^*}}(\beta_n)\to0\), contradicting
\(L(\beta_n)\ge\varepsilon\). Thus \(L(\beta)\to0\) as
\(\|\beta\|\to\infty\).

Now take any finite \(\beta_{\rm ref}\in\calX\). Since \(L(\beta_{\rm ref})>0\),
there exists \(R<\infty\) such that \(L(\beta)<L(\beta_{\rm ref})\) whenever
\(\|\beta\|\ge R\). Therefore the supremum over \(\calX\) is attained on the
compact set \(\{\beta\in\calX:\|\beta\|\le R\}\), by continuity of \(L\). Hence
a finite maximizer exists.
\end{proof}

While the rapidly varying models cover many standard links, this class is not exhaustive. There exist popular heavy-tailed link functions, most notably the Cauchit link, that do not fulfill the rapid variation assumptions. The asymptotic behavior of these models is fundamentally different and requires a separate analysis. We investigate these models now.

\section{Regular variation}\label{sec:regular}

We next consider the regularly varying tail class.

\begin{definition}[Regular variation at $+\infty$]
Let $\tau:(0,\infty)\to(0,\infty)$ be ultimately increasing. We say that $\tau$ is regularly varying at $+\infty$ with regular-variation index $\rho_r \in (0, \infty)$ if for any $a, b > 0$,
\begin{equation}
\lim_{t\to\infty} R_{ab}(t) = \left(\frac{b}{a}\right)^{\rho_r}.
\label{eq:RRV-scale}
\end{equation}
The characterization theorem \citep[Thm~1.4.1]{bingham1989regular} gives the equivalent form
\[
\tau(t) = t^{\rho_r} L(t)
\]
where $L(t)$ is a slowly varying function, meaning $\lim_{t\to\infty} L(xt)/L(t) = 1$ for any $x>0$. This implies that for $\tau(s) = s^{\rho_r} L(s)$,
\[
\lim_{s\to\infty} \frac{\tau(as)}{\tau(s)} = \lim_{s\to\infty} \frac{(as)^{\rho_r} L(as)}{s^{\rho_r} L(s)} = a^{\rho_r} \lim_{s\to\infty} \frac{L(as)}{L(s)} = a^{\rho_r}.
\]
\end{definition}

\begin{definition}[Regularly varying model]
A model in this family is regularly varying if its odds transform $\tau$ is regularly varying at $+\infty$ with right-tail regular-variation index ${\rho_r} > 0$ and its left-tail transform $\tilde{\tau}(s) = 1/\tau(-s)$ is regularly varying at $+\infty$ with left-tail regular-variation index ${\rho_l} > 0$.
\end{definition}

\begin{lemma}[Asymptotic ratios for regularly varying models]\label{lem:rr-ratio-limit}
For a regularly varying model with right-tail index ${\rho_r} > 0$ and left-tail index ${\rho_l} > 0$, let $a, b \in \mathbb{R}$ satisfy $a > b$, and let $c,d\in\R$. The limit of the ratio $R(t) = \frac{\tau(b t + d)}{\tau(a t + c)}$ as $t \to \infty$ is as follows.
\begin{enumerate}
    \item If $a > b > 0$, then $\lim_{t\to\infty} R(t) = \left(\frac{b}{a}\right)^{\rho_r} \in (0, 1)$.
    \item If $0 > a > b$, then $\lim_{t\to\infty} R(t) = \left(\frac{|a|}{|b|}\right)^{\rho_l} \in (0, 1)$.
    \item If $a > 0 \ge b$ or $a = 0, b < 0$, then $\lim_{t\to\infty} R(t) = 0$.
\end{enumerate}
\end{lemma}

\begin{lemma}[Ties for regularly varying models]\label{lem:rv-ties}
Let $\tau$ be the odds transform of a regularly varying model with indices ${\rho_r}, \rho_l>0$.
Fix $A\neq 0$ and bounded $c,d\in\R$. Then
\[
\frac{\tau(A t + d)}{\tau(A t + c)}
\,\xrightarrow[t\to\infty]{}\, 1.
\]
\end{lemma}

\begin{proposition}[Affine-ray completion of the boundary]
\label{prop:rr-ray-completion}
For a regularly varying fixed-reference model on the finite-dimensional
estimable space \(\calX\), every subsequential limit of \(L(\beta_n)\) along
a sequence satisfying \(\|\beta_n\|\to\infty\) belongs to the closure of
\[
\left\{
L_\infty(\beta_0,v):
\beta_0\in\calX,\;
v\in\calX\setminus\{0\}
\right\}.
\]
Consequently, when $\dim\mathcal X\ge1$,
\[
 L_{\partial}
 =\lim_{R\to\infty}
      \sup_{\substack{\beta\in\mathcal X\\\|\beta\|\ge R}}L(\beta)
 =\max_{\substack{v\in\mathcal X\\\|v\|=1}}
      \sup_{\beta_0\in\mathcal X}L_\infty(\beta_0,v).
\]
The same identity holds for the log-likelihood, with
\(\log 0=-\infty\).
\end{proposition}

The proof, given in the Supplementary Material, compares a divergent sequence $\beta_n$ with the affine ray that starts at $\beta_n$ itself and points in the limiting direction $v$ of $\beta_n/\|\beta_n\|$. Under regular variation the two likelihoods differ by $o(1)$, so curved or multiscale escape paths cannot raise the boundary supremum, and a maximizing direction always exists.

The primary example of a regularly varying model is the family of Student's t-links. The probability density function satisfies $f_\nu(s) \propto (1+s^2/\nu)^{-(\nu+1)/2}$, which implies the asymptotic tail behavior
\[
f_\nu(s) \sim c_\nu s^{-(\nu+1)} \qquad (s \to \infty)
\]
for a constant $c_\nu>0$. The survival function behaves as the integral of the density tail
\[
\bar{F}_\nu(t) = \int_t^\infty f_\nu(s)\,ds \sim c_\nu\int_t^\infty s^{-(\nu+1)}\,ds = \frac{c_\nu}{\nu}\,t^{-\nu},
\]
and thus the odds relationship $\tau(s) \sim 1/\bar{F}_\nu(s)$ implies
\[
\tau(s) \sim c \cdot s^{\nu}.
\]
This is a regularly varying function with index ${\rho_r} = \nu$. By symmetry, the left-tail function $\tilde{\tau}(s)$ also has index ${\rho_l} = \nu$. Therefore, a t-link with $\nu$ degrees of freedom is a regularly varying model with ${\rho_r} = {\rho_l} = \nu$.

The Cauchit link is the simplest member of this family, as it is a t-link with $\nu=1$ degree of freedom. For the Cauchit model, we thus have ${\rho_r} = 1$.

Writing $\tau(s)=s^{\rho_r} L(s)$, the logarithmic derivative is $\lodds'(s) = {\rho_r}/s + L'(s)/L(s)$. If $\tau$ is continuously differentiable in the right tail and $\tau'$ is ultimately monotone, the monotone density theorem \citep[Thm~1.7.2]{bingham1989regular} applies to the asymptotic relation $\tau(s) \sim s^{\rho_r} L(s)$ to obtain $\tau'(s) \sim {\rho_r} s^{{\rho_r}-1} L(s)$. This yields
\[
\lodds'(s) = \frac{\tau'(s)}{\tau(s)} \sim \frac{{\rho_r} s^{{\rho_r}-1} L(s)}{s^{\rho_r} L(s)} = \frac{{\rho_r}}{s} \qquad (s \to \infty).
\]
Under this assumption the right-tail log-odds derivative vanishes.

\subsection{Escape geometry}

We now define the geometric conditions that govern the likelihood's asymptotic behavior. Let $j=y_i$ be the observed category for observation $i$. As in Definition~\ref{def:span-sep}, every ray satisfies $v\in\calX\setminus\{0\}$.

\begin{definition}[S1, quasi-one-vs-all separation] \label{S1}
The structure set exhibits quasi-one-vs-all separation if there exists $v \in\calX\setminus\{0\}$ such that for every observation $i$ (with $j=y_i$):
\[
(z_{ij}^\T v \ge 0) \land (z_{ik}^\T v \le 0, \forall k \ne j).
\]
\end{definition}

\begin{definition}[S2, quasi-complete pairwise separation] \label{S2}
The structure set exhibits quasi-complete pairwise separation if there exists $v \in\calX\setminus\{0\}$ such that for every observation $i$ (with $j=y_i$):
\[
(z_{ij} - z_{ik})^\T v \ge 0, \quad \forall k \ne j,
\]
and at least one of these inequalities is strict for some pair \((i,k)\). 
\end{definition}

\begin{definition}[ESC, Escape condition] \label{ESC}
The structure set exhibits an escape condition if there exists $v \in\calX\setminus\{0\}$ such that for every observation $i$ (with $j=y_i$), one of the three scenarios holds
\[
(z_{ij}^\T v > 0) \quad \lor \quad (\max_{k \in [J]} z_{ik}^\T v < 0) \quad \lor \quad (\max_{k \in [J]} z_{ik}^\T v = 0 = z_{ij}^\T v).
\]
\end{definition}
With a fixed reference, $z_{iJ}=0$ excludes the second alternative, which is kept for general category scores.

\begin{lemma}[Separation hierarchy]\label{lem:hierarchy}
The geometric conditions are nested as follows.
\[
S1 \implies S2 \implies ESC.
\]
\end{lemma}

\begin{definition}[Anti-escape]
The structure set exhibits an anti-escape condition if it does not exhibit an escape condition. This holds if for every $v \in\calX\setminus\{0\}$ there exists at least one observation $i$ (with $j=y_i$) such that
\[
\left(\max_{k \in [J]} z_{ik}^\T v > 0 \ge z_{ij}^\T v \right) \quad \lor \quad \left(\max_{k \in [J]} z_{ik}^\T v = 0 > z_{ij}^\T v \right).
\]
\end{definition}
In the anti-escape condition, any positive or zero row maximum relevant for anti-escape must be attained by a competitor, so the formulation is equivalent to restricting the maximum to $[J]\setminus\{j\}$.

These conditions are computationally checkable. After choosing a basis of the estimable span $\mathcal X$ and writing $v=Bu$, the conditions in Definitions~\ref{S1} and~\ref{S2} are feasibility problems for a single polyhedral cone. For the escape condition, fixing for each observation which of the three alternatives in Definition~\ref{ESC} holds again gives linear sign restrictions. Hence escape is feasibility over a finite union of polyhedral cones, or equivalently one mixed-integer linear feasibility problem. Anti-escape means that no such nonzero direction exists. Thus, for regularly varying links, \(S1\) gives nonexistence, anti-escape gives existence, and the only nondecisive case is escape without \(S1\). In that plateau regime, Proposition~\ref{prop:rr-ray-completion} shows that the full boundary supremum is obtained by optimizing the affine-ray limits over both the direction \(v\) and the finite starting point \(\beta_0\). Geometry alone does not determine whether this boundary value exceeds an interior likelihood value.

\subsection{Existence of the maximum likelihood estimator}

We now connect the geometric separation hierarchy to the asymptotic behavior of the likelihood and the existence of the MLE.
For any starting point $\beta_0$ and ray $v$, the limit of the likelihood $L_\infty(\beta_0,v) = \lim_{t\to\infty} L(\beta_0 + tv)$ is the product of the limits of the individual probabilities, $L_\infty(\beta_0,v) = \prod_i c_i(\beta_0,v)$, where
\[
c_i(\beta_0,v) = \lim_{t\to\infty} h_{i,y_i}(\beta(t)) = \lim_{t\to\infty} \frac{1}{1 + \sum_{k \ne y_i} R_{ik}(t)}.
\]
The behavior of $c_i(\beta_0,v)$ is determined by the limits of the ratios $R_{ik}(t) = \tau(a_{ik}t + b_{ik}) / \tau(a_{ij}t + b_{ij})$. Lemma \ref{lem:rr-ratio-limit} and its inverse give the criterion below.

A ratio $R_{ik}(t)$ will diverge to infinity, causing $c_i(\beta_0,v) \to 0$, if and only if the slopes satisfy either $(a_{ik} > 0 \ge a_{ij})$ or $(a_{ik} = 0 > a_{ij})$. 

In all other cases, the ratio $R_{ik}(t)$ converges to a finite constant. The constant is zero, one, a value in $(0,1)$, or a value in $(1,\infty)$ depending on the relationship between $a_{ik}$ and $a_{ij}$ as detailed in Lemma \ref{lem:rr-ratio-limit}.

The value $c_i(\beta_0,v)$ can only be 1 if $1 + \sum \lim R_{ik}(t) = 1$, which requires $\lim R_{ik}(t) = 0$ for all $k \ne j$. Based on our lemma, the only way for the limit to be 0 is if $a_{ij} > a_{ik}$ and either $a_{ij} > 0 \ge a_{ik}$ or $a_{ij}=0, a_{ik}<0$. For this to hold for all competitors $k$, either $a_{ij} > 0$ and $a_{ik}\le 0$ for all $k\ne j$, or $a_{ij}=0$ and $a_{ik}<0$ for all $k\ne j$. The second alternative is the fixed-reference case in which the observed category has slope zero. Together these alternatives are the strict version of the S1 condition.

\begin{theorem}[Regular variation asymptotic trichotomy]\label{thm:rr-main}
Let $\calS$ be defined for a regularly varying model on $\calX$. The asymptotic regimes along rays, and the resulting existence conclusions in the decisive cases, are as follows.
\begin{enumerate}
    \item[(a)] If quasi-one-vs-all separation holds, then no finite MLE exists.
    
    \item[(b)] If the escape condition holds but quasi-one-vs-all separation fails, then every such escape ray has a positive subunit likelihood limit. This identifies the intermediate plateau regime. By Proposition~\ref{prop:rr-ray-completion}, the full boundary supremum is the supremum of these affine-ray limits over starting points and directions. It may or may not dominate likelihood values attained in the interior.
    
    \item[(c)] If anti-escape holds, then $L_\infty(\beta_0,v) = 0$ for all finite $\beta_0$ and all rays $v\in\calX\setminus\{0\}$, and a finite MLE exists.
\end{enumerate}
\end{theorem}
\begin{proof}
Fix a ray \(v\in\calX\setminus\{0\}\) and a starting point \(\beta_0\), and write
\(\beta(t)=\beta_0+tv\). For observation \(i\), let \(j=y_i\), and define
\[
a_{ik}=z_{ik}^{\T}v,\qquad b_{ik}=z_{ik}^{\T}\beta_0,\qquad
R_{ik}(t)=
\frac{\tau(a_{ik}t+b_{ik})}{\tau(a_{ij}t+b_{ij})}.
\]
Then \(L_\infty(\beta_0,v)=\prod_i c_i(\beta_0,v)\), where
\[
c_i(\beta_0,v)
=
\lim_{t\to\infty}
\frac{1}{1+\sum_{k\ne j}R_{ik}(t)}.
\]

\begin{enumerate}
\item[(a)] Suppose \(S1\) holds. Then \(a_{ij}\ge0\) and \(a_{ik}\le0\) for all
\(i\) and all \(k\ne j\). Since \(v\ne0\) in the estimable span, not all slopes
can vanish, and therefore at least one row has a strict inequality. In that row,
at least one competitor ratio tends to zero by Lemma~\ref{lem:rr-ratio-limit},
and no competitor ratio diverges. All other rows have limiting denominators no
larger than their initial denominators. Hence
\[
\lim_{t\to\infty}L(\beta_0+tv)>L(\beta_0).
\]
Since this holds from every finite \(\beta_0\), no finite point can maximize the
likelihood. Thus no finite MLE exists.

\item[(b)] Let \(v\) be an escape ray that does not satisfy \(S1\). Under escape,
each row satisfies one of
\[
a_{ij}>0,\qquad \max_k a_{ik}<0,\qquad
a_{ij}=0=\max_k a_{ik}.
\]
In all three cases Lemmas~\ref{lem:rr-ratio-limit} and~\ref{lem:rv-ties} show that no ratio
\(R_{ik}(t)\) diverges, so \(c_i(\beta_0,v)\in(0,1]\) and
\(L_\infty(\beta_0,v)>0\).

Since \(v\) does not satisfy \(S1\), at least one row violates the \(S1\)
inequalities. Under escape this can only happen in one of two ways. Either
\(a_{ij}>0\) and some competitor has \(a_{ik}>0\), in which case
\(R_{ik}(t)\) converges to a positive finite constant, or \(\max_k a_{ik}<0\),
in which case all row slopes are negative and all ratios converge to positive
finite constants. In either case \(c_i(\beta_0,v)\in(0,1)\) for at least one
row. Hence
\[
0<L_\infty(\beta_0,v)<1.
\]
This is the intermediate plateau regime. In this regime separation geometry
alone is not decisive. Finite existence depends on the comparison between
boundary limits and interior likelihood values.

\item[(c)] Assume anti-escape. Fix \(\beta_0\in\calX\) and
\(v\in\calX\setminus\{0\}\). By anti-escape, there is an observation
\(i^*\), with \(j=y_{i^*}\), and a competitor \(k^*\) attaining the row
maximum such that either
\[
z_{i^*k^*}^\T v>0\geq z_{i^*j}^\T v,
\qquad\text{or}\qquad
z_{i^*k^*}^\T v=0>z_{i^*j}^\T v.
\]
Lemma~\ref{lem:rr-ratio-limit} therefore gives
\[
\frac{\tau\{z_{i^*j}^\T(\beta_0+tv)\}}
     {\tau\{z_{i^*k^*}^\T(\beta_0+tv)\}}
\longrightarrow0.
\]
Consequently,
\[
h_{i^*,j}(\beta_0+tv)
\leq
\frac{\tau\{z_{i^*j}^\T(\beta_0+tv)\}}
     {\tau\{z_{i^*k^*}^\T(\beta_0+tv)\}}
\longrightarrow0,
\]
and hence \(L_\infty(\beta_0,v)=0\). Since this holds for every affine ray,
Proposition~\ref{prop:rr-ray-completion} gives \(L_\partial=0\), equivalently
\[
L(\beta)\longrightarrow0
\qquad\text{as }\|\beta\|\to\infty\text{ in }\calX.
\]
Choose any finite \(\beta_{\rm ref}\in\calX\). Since
\(L(\beta_{\rm ref})>0\), there exists \(R<\infty\) such that
\(L(\beta)<L(\beta_{\rm ref})\) whenever \(\|\beta\|\geq R\). The supremum is
therefore attained on the compact ball
\(\{\beta\in\calX:\|\beta\|\leq R\}\), by continuity of \(L\). Hence a
finite maximizer exists.\qedhere{}
\end{enumerate}
\end{proof}

The distinction between pairwise overlap and anti-escape is essential: pairwise overlap does not preclude an escape direction, and therefore does not ensure existence under regular variation.

\begin{proposition}[Overlap and tail-index reversal]\label{prop:reversal}
Let $n=p=1$, $J=3$, $y_1=1$ and $(z_{11},z_{12},z_{13})=(2,3,0)$, and consider the fixed-reference model with odds transform $\tau_\rho(s)=\exp\{\rho\,\mathrm{arsinh}(s)\}$, $\rho>0$. The structure set exhibits pairwise overlap. No finite maximum likelihood estimator exists when $\rho=1$, whereas a finite maximum likelihood estimator exists when $\rho=2$.
\end{proposition}

\begin{proof}
Since $\tau_\rho(s)=\{s+(1+s^2)^{1/2}\}^\rho$ is strictly increasing with $\tau_\rho(0)=1$, $\tau_\rho(-s)=1/\tau_\rho(s)$ and $\tau_\rho(s)\sim(2s)^\rho$ as $s\to\infty$, it is the centered odds transform of the distribution function $F_\rho=\tau_\rho/(1+\tau_\rho)$, and its right-tail odds and left-tail transform are both regularly varying with index $\rho$. The structure set is $\mathcal S=\{z_{11}-z_{12},z_{11}-z_{13}\}=\{-1,2\}$, so no nonzero $v$ makes both structure inequalities nonnegative, and the data exhibit pairwise overlap. The direction $v=1$ is nevertheless an escape direction, since $z_{11}v=2>0$, whereas quasi-one-vs-all separation fails: $v>0$ gives the competitor slope $3v>0$, and $v<0$ gives the observed slope $2v<0$. The data therefore lie in the plateau regime of Theorem~\ref{thm:rr-main}(b). The likelihood is
\[
L_\rho(\beta)=\frac{\tau_\rho(2\beta)}{1+\tau_\rho(2\beta)+\tau_\rho(3\beta)},
\]
and regular variation gives $L_\rho(\beta)\to2^\rho/(2^\rho+3^\rho)$ as $\beta\to\infty$ and $L_\rho(\beta)\to0$ as $\beta\to-\infty$.

For $\rho=1$, the inequality $L_1(\beta)<2/5$ is equivalent to $3\tau_1(2\beta)<2\tau_1(3\beta)+2$. After the linear terms cancel, this becomes $3(1+4\beta^2)^{1/2}<2(1+9\beta^2)^{1/2}+2$. Both sides are positive, and the difference of their squares is $8(1+9\beta^2)^{1/2}-1>0$, so the inequality holds for every finite $\beta$. Hence $\sup_\beta L_1(\beta)=2/5$ is not attained.

For $\rho=2$, $L_2(0)=1/3$ exceeds both boundary limits, $4/13$ and $0$. By continuity, $L_2(\beta)<L_2(0)$ outside some compact interval, on which the maximum is attained.
\end{proof}

Thus the same data and the same pairwise geometry yield opposite existence conclusions when only the regular-variation index changes.

Proposition~\ref{prop:rr-ray-completion}  also simplifies practical analysis of the plateau regime. The full boundary supremum may be obtained from affine rays, so no separate multiscale analysis of curved paths is required. A general algorithm for computing and certifying this supremum is left for future work.

\section{Uniqueness}

We now study global concavity in the full normalized-coordinate map. Throughout this section assume $\tau>0$, $\tau\in C^2$, $\lodds=\log\tau$, $\lodds(s)\to\infty$ as $s\to\infty$ and $\lodds(s)\to-\infty$ as $s\to-\infty$. These assumptions hold for centered odds transforms generated by strictly increasing, twice continuously differentiable distribution functions. As before, 
\[
  h_j(\gamma)=\frac{\tau(\gamma_j)}{\sum_{k=1}^J \tau(\gamma_k)},
  \qquad
  \gamma=(\gamma_1,\dots,\gamma_J)\in\mathbb{R}^J,
  \qquad
  \lodds=\log\tau.
\]
The fixed-reference model used above is obtained by restricting this map to the affine slice $\gamma_J=0$. On the binary fixed-reference slice, the probit likelihood can be strongly log-concave under the geometric conditions of \citet{chak-rico-zanella-2026}. The result below concerns global concavity of the full normalized map for arbitrary alternative-specific linear predictors.

For observation $i$, write
\[
  \gamma_i=(\gamma_{i1},\dots,\gamma_{iJ}),
  \qquad
  \gamma_{ik}=z_{ik}^\T\beta,
  \qquad
  h_{ij}(\beta)=h_j(\gamma_i).
\]
Define the componentwise transformed utilities $u_{ik}=\lodds(\gamma_{ik})$ and the softmax probabilities
\[
p_{ik} = h_{ik}(\beta)=h_k(\gamma_i)= \frac{e^{u_{ik}}}{\sum_{m=1}^J e^{u_{im}}}.
\]
Let the multinomial-logit objective in $u$ be
\[
\Gamma(u) = u_{y_i}-\log\!\Big(\sum_{k=1}^J e^{u_k}\Big),
\]
which is globally concave in $u\in\mathbb R^J$. The per-observation log-likelihood as a function of $\gamma_i$ is the composition
\[
\loglik_i(\gamma_i) = \Gamma\big(u_i\big) = \Gamma\big(\lodds(\gamma_i)\big).
\]

\subsection{Criterion for global concavity}

Let $\delta_{rs}$ be the Kronecker delta function. The composition $\loglik_i = \Gamma(\lodds(\gamma_i))$ immediately reveals a deep connection between the model form and its concavity. This is formalized in the following lemma and theorem, which show that global concavity is a unique property of the logit model.

\begin{lemma}
\label{lem:hessian-form}
Let $\loglik_i(\gamma_i)$ be the per-observation log-likelihood, and for fixed $i$ suppress the first index by writing $\gamma_r=\gamma_{ir}$ and $p_r=h_r(\gamma_i)=h_{ir}(\beta)$. The entries of the Hessian matrix $\nabla^2 \loglik_i(\gamma_i)$ are given by
\[
\frac{\partial^2 \loglik_i}{\partial \gamma_r \partial \gamma_s}
=\underbrace{\delta_{rs} \lodds''(\gamma_r) (\delta_{ry_i} - p_r)}_{\text{diagonal correction}}
\underbrace{- \lodds'(\gamma_r) \lodds'(\gamma_s) p_r (\delta_{rs} - p_s)}_{\text{variance term}}.
\]
Furthermore, for any vector $a \in \mathbb{R}^J$, the quadratic form of the variance term is
\[
\sum_{r=1}^J \sum_{s=1}^J a_r \left[ - \lodds'(\gamma_r) \lodds'(\gamma_s) p_r (\delta_{rs} - p_s) \right] a_s = - \mathrm{Var}_{p_i}(R),
\]
where $R$ is a discrete random variable taking value $w_k = \lodds'(\gamma_k) a_k$ with probability $p_k$.
\end{lemma}

\begin{theorem}
\label{thm:expfam-iff}
Let $\tau:\mathbb{R}\to(0,\infty)$ be twice continuously differentiable. Let $\lodds=\log\tau$ satisfy $\lodds(s)\to\infty$ as $s\to\infty$ and $\lodds(s)\to-\infty$ as $s\to-\infty$. Consider the full normalized-coordinate reference-ratio model
\[
h_j(\gamma)=\frac{\tau(\gamma_j)}{\sum_{k=1}^J \tau(\gamma_k)},
\qquad\text{with}\qquad \gamma = Z^\T\beta,
\]
and the total log-likelihood $\loglik(\beta)=\sum_{i=1}^n\big(\lodds(\gamma_{i,y_i})-\log\sum_{k=1}^J e^{\lodds(\gamma_{ik})}\big)$.
The following statements are equivalent.
\begin{enumerate}
\item[(a)] $\loglik(\beta)$ is globally concave in $\beta$ for every design $Z$ and every response $y$.
\item[(b)] $\lodds$ is affine, i.e., $\lodds(u)=a u+b$ for some $a> 0$, $b\in\R$.
\item[(c)] The family $\{h(\gamma):\gamma\in\R^J\}$ forms a regular exponential family in the canonical parameter $\eta=a\gamma$, \(a>0\), with cumulant $K(\eta)=\log\sum_{k=1}^J e^{\eta_k}$, i.e.,
\[
h_j(\gamma)=\frac{e^{a\gamma_j}}{\sum_k e^{a\gamma_k}}
\quad\text{and}\quad
\loglik(\beta)=\sum_{i=1}^n\big(\eta_{i,y_i}-K(\eta_i)\big) + \text{const}.
\]
\end{enumerate}
\end{theorem}

\begin{proof}
\noindent
\emph{(b) $\implies$ (c).} If $\lodds(u)=au+b$, then
\[
h_j(\gamma)=
\frac{e^{a\gamma_j+b}}{\sum_k e^{a\gamma_k+b}}
=
\frac{e^{a\gamma_j}}{\sum_k e^{a\gamma_k}}.
\]
With $\eta=a\gamma$ and $K(\eta)=\log\sum_k e^{\eta_k}$, the per-observation contribution is $\eta_{y_i}-K(\eta)$. This is the canonical multinomial exponential-family form.

\noindent
\emph{(c) $\implies$ (a).} The cumulant $K$ is convex, so $\eta_{y_i}-K(\eta)$ is concave in $\eta$. Since $\eta_i=aZ_i^\T\beta$ is affine in $\beta$, the total log-likelihood is concave in $\beta$.

\noindent
\emph{(a) $\implies$ (b).} Global concavity for every design implies concavity of each per-observation log-likelihood in the full coordinate vector $\gamma$. By Lemma~\ref{lem:hessian-form}, for an observation with response category $y_i$,
\[
H_{rs}
=
\delta_{rs}\lodds''(\gamma_r)(\delta_{ry_i}-p_r)
-\lodds'(\gamma_r)\lodds'(\gamma_s)p_r(\delta_{rs}-p_s).
\]
For $J>2$, fix $y_i=1$ and send $\gamma_k\to-\infty$ for $k=3,\ldots,J$. Then $p_k\to0$, and the $\{1,2\}$ principal Hessian block converges to the two-category block. Since principal submatrices and limits of negative semidefinite matrices are negative semidefinite, it is enough to analyze $J=2$. Let $p=p_1$ and $p_2=1-p$. If $\gamma_1=s$ is fixed and $\gamma_2\to+\infty$, then $p\to0$ and the limiting Hessian has diagonal entry $\lodds''(s)$, so negative semidefiniteness requires $\lodds''(s)\le0$. If instead $\gamma_2=s$ is fixed and $\gamma_1\to-\infty$, then $p\to0$ and the limiting Hessian has diagonal entry $-\lodds''(s)$, so negative semidefiniteness requires $\lodds''(s)\ge0$. Since $s$ is arbitrary, $\lodds''\equiv0$, and $\lodds$ is affine. Since \(\lodds(s)\to\infty\) as \(s\to\infty\) and
\(\lodds(s)\to-\infty\) as \(s\to-\infty\), the affine slope must be positive. Thus \(\lodds(u)=au+b\) with \(a>0\).
\end{proof}

\begin{corollary}
\label{cor:logit-unique}
Among all reference-ratio links $h_j(\gamma)=\tau(\gamma_j)/\sum_k\tau(\gamma_k)$,
global concavity of \(\ell(\beta)\) for all designs holds if and only if \(\tau(u)=Ce^{au}\) for some \(C>0\) and \(a>0\).
Equivalently, the multinomial logit (standard softmax) is the unique globally concave case (up to linear rescaling).
\end{corollary}

\subsection{Fixed-reference concavity and tail implications}

While Theorem~\ref{thm:expfam-iff} uses non-strict concavity to characterize the unrestricted normalized-coordinate map, the resulting multinomial logit log-likelihood is strictly concave on the estimable span under the usual identifiability condition. Concavity on the fixed-reference slice \(\gamma_J=0\) is a weaker property and, for a fixed finite number of categories, need not force affine log odds.

\begin{proposition}\label{prop:fixedref}
Let \(m=J-1\ge2\), and let \(\tau\) be twice continuously differentiable with \(\tau'(s)>0\) for every \(s\in\R\) and \(\tau(s)\to\infty\) as \(s\to\infty\). Then the fixed-reference log-likelihood is globally concave in the regression parameter for every design and every response vector if and only if
\begin{equation}
-\frac{\{\lodds'(s)\}^{2}}{1+m\tau(s)}
\leq
\lodds''(s)
\leq 0
\qquad (s\in\R).
\label{eq:fixed-reference-concavity}
\end{equation}
\end{proposition}
Its proof is given in the Supplementary Material. This finite-dimensional qualification does not alter the tail-concavity tradeoff relevant here. The displayed inequalities imply that
\[
\lodds'(s)\longrightarrow a_+\in(0,\infty),
\qquad
\lodds(s)=a_+s+o(s),
\qquad s\to\infty.
\]
Thus, for $J\ge3$, a globally concave fixed-reference link has asymptotically linear log odds in the right tail, which rules out regularly varying links such as Cauchit and Student $t$. Moreover, if the same link is required to yield a globally concave fixed-reference likelihood uniformly over the number of categories, then letting \(m\to\infty\) in the lower bound forces \(\lodds''\ge0\). Together with the upper bound \(\lodds''\le0\), this gives \(\lodds''\equiv0\). Multinomial logit is therefore the unique globally concave member of the unrestricted normalized-coordinate family and the unique dimension-uniform globally concave fixed-reference link. Heavy-tailed links gain their separation robustness by leaving this dimension-stable globally concave class.

Table~\ref{tab:link-boundary-summary} summarizes the right-tail ratio limits, asymptotic sensitivity and global concavity of the links considered here.

\begin{table}[tb]
\centering
\caption{Boundary behaviour of common reference-ratio links. Let
\(R_{ab}(t)=\tau(bt)/\tau(at)\) for \(a>b>0\). The tail-sensitivity column gives
the leading behaviour of \(\lodds'(s)\), where \(\lodds(s)=\log\tau(s)\), as
\(s\to\infty\). The last column records global concavity for every design in the full
normalized-coordinate formulation.}
\label{tab:link-boundary-summary}
\renewcommand{\arraystretch}{1.18}
\begin{tabular}{@{}llllc@{}}
\hline
Link & Tail class & Strict advantage \(R_{ab}(t)\) & Tail sensitivity \(\lodds'(s)\) & Concave \\
\hline
Logit
& rapid
& \(\exp\{-(a-b)t\}\)
& \(1\)
& \(\checkmark\) \\

Probit
& rapid
& \(\exp\{-(a^2-b^2)t^2/2+o(t^2)\}\)
& \(s+o(s)\)
& \(\times\) \\

Cloglog
& rapid
& \(\exp\{c_0(e^{bt}-e^{at})+o(1)\}\)
& \(c_0e^s+o(e^s)\)
& \(\times\) \\

Cauchit
& regular, \(\rho_r=1\)
& \((b/a)+o(1)\)
& \(1/s+o(1/s)\)
& \(\times\) \\

Student-\(t_\nu\)
& regular, \(\rho_r=\nu\)
& \((b/a)^\nu+o(1)\)
& \(\nu/s+o(1/s)\)
& \(\times\) \\
\hline
\end{tabular}

\medskip
\begin{minipage}{0.90\textwidth}
\footnotesize
The same-tail case \(a>b>0\) is the case that distinguishes rapid and regular
tails most sharply. Rapid tails eliminate a strict competitor asymptotically,
whereas regular tails leave a positive boundary contribution. The sensitivity
\(\lodds'(s)\) explains tied-slope behaviour: finite shifts remain visible for
logit, are amplified in the right tails of probit and complementary log-log, and are erased asymptotically
for regularly varying odds. If the advantage crosses tails, for example
\(a>0\ge b\), then \(R_{ab}(t)\to0\) for both tail classes.
\end{minipage}
\end{table}

\section{Numerical illustration}\label{sec:numerical}

We illustrate the intermediate plateau regime with a fixed-reference example with \(J=3\), \(z_{i3}=0\), \(n=8\), and \(y_i=1\) for all \(i\). The two moving category scores are
\[
z_{11}=(6/5,0),\quad z_{12}=(1/5,0),\qquad
z_{i1}=(2,0),\quad z_{i2}=(1,3)\quad (i=2,\ldots,7),
\]
and \(z_{81}=(1/20,1)\), \(z_{82}=(1/20,0)\). 

The pairwise structure vectors $s_i = z_{i1} - z_{i2}$ against category \(2\) are \((1,0)\), \((1,-3)\) repeated six times, and \((0,1)\). The structure vectors against the fixed reference category are \(z_{i1}\). Hence complete pairwise separation holds whenever $v_1>0$, $v_2>0$ and $v_1-3v_2>0$.

\begin{figure}[htbp]
\centering
\includegraphics[width=0.90\textwidth]{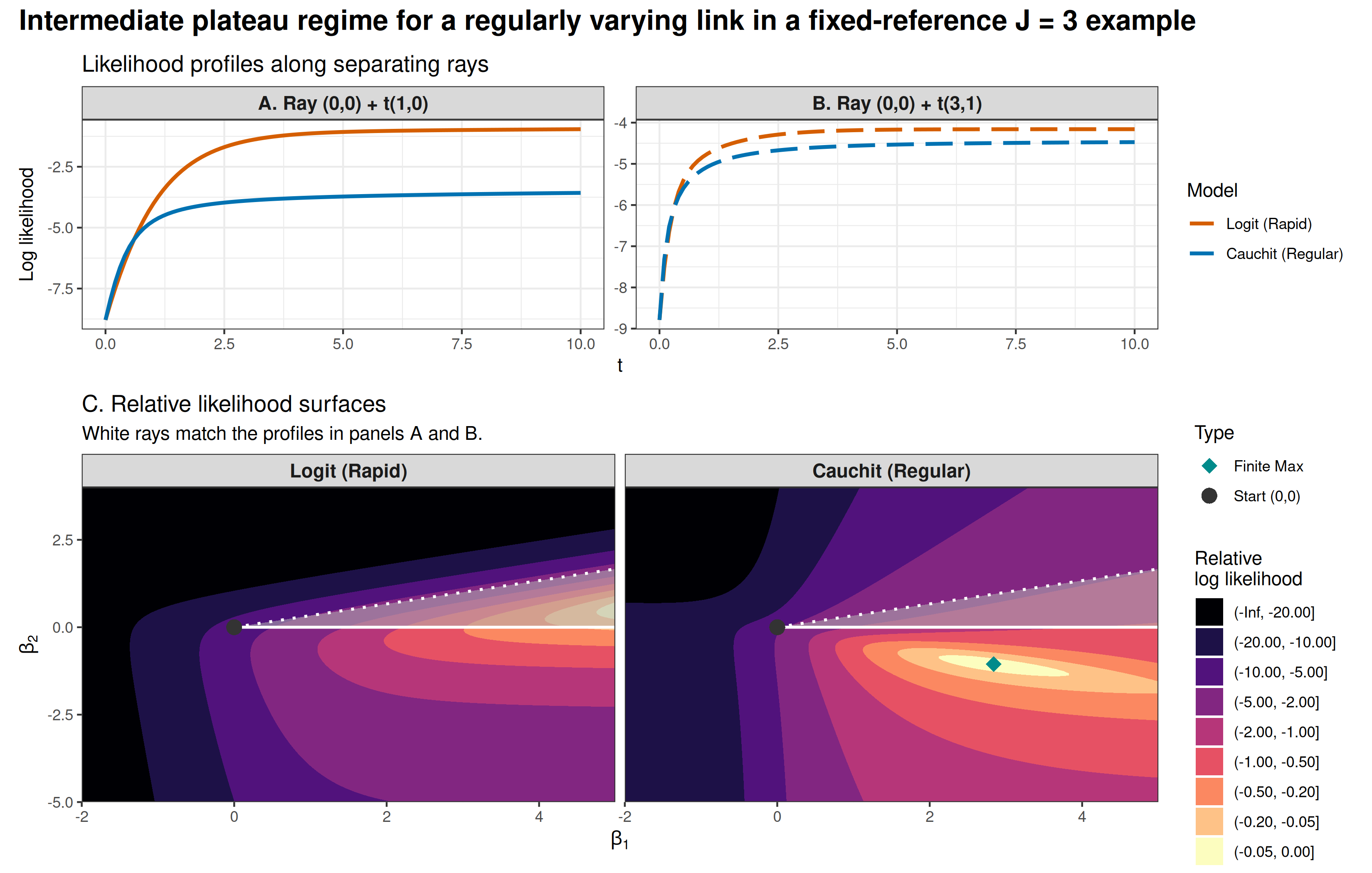}
\setlength{\figwidth}{\textwidth}
\caption{Intermediate plateau regime in a fixed-reference \(J=3\) model. The upper panels show likelihood profiles along the boundary rays \((1,0)\) and \((3,1)\). Logit has boundary plateaux on these rays, while strict interior separating rays approach the unattained supremum. Cauchit converges to lower plateaux. The lower panels show relative log-likelihood contours, the separating cone and the finite Cauchit maximizer.}
\label{fig:plt_sim}
\alttext{Boundary profiles and contour plots compare logit and Cauchit likelihoods in a fixed-reference three-category example. Logit estimates diverge along strict separating rays, while the Cauchit likelihood levels off and has a finite maximum outside the separating cone.}
\end{figure}

The closure of this strict system defines the polyhedral cone $C = \{v \in \mathbb{R}^2 \mid v_1 \ge 3v_2 \ge 0\}$, shown in Fig.~\ref{fig:plt_sim}. After normalizing \(v=(1,r)\), the strict system holds exactly for \(0<r<1/3\), the interior of \(C\). Proposition~\ref{prop:complete-sep} therefore implies that the logit model has no finite maximum likelihood estimator.

For the Cauchit model, \(S1\) (Definition~\ref{S1}) is infeasible. The first observation forces \(v_1\ge0\) and \(v_1\le0\), hence \(v_1=0\). The middle observations then force \(v_2\le0\), while the last observation forces \(v_2\ge0\). Thus only \(v=0\) satisfies the \(S1\) inequalities. Escape directions nevertheless exist. Every nonzero escape direction has \(v_1>0\), so write \(v=(1,r)\). The last row requires \(r+1/20>0\), and hence the escape directions are exactly \(r>-1/20\).

For such a ray, regular variation with Cauchit right-tail index \(\rho_r=1\) gives the limiting row contributions
\[
c_1=\frac67,\qquad
c_i=\frac{2}{3(1+r)}\quad (i=2,\ldots,7),\qquad
c_8=\frac{r+1/20}{r+1/10}.
\]
Therefore
\[
\loglik_\infty(r)
=
\log\frac67
+
6\log\left\{\frac{2}{3(1+r)}\right\}
+
\log\left(\frac{r+1/20}{r+1/10}\right), \qquad r>-1/20.
\]
A derivative calculation gives the maximizing escape direction
\[
r_*=\frac{\sqrt{481}-17}{240}\approx0.0205,
\qquad
\loglik_\infty(r_*)\approx -3.245.
\]
Non-escape rays have likelihood limit zero. By Proposition~\ref{prop:rr-ray-completion}, affine-ray limits exhaust the full boundary supremum, so maximizing the displayed envelope gives
\[
\lim_{R\to\infty}
\sup_{\substack{\beta\in\calX\\\|\beta\|\geq R}}
\loglik(\beta)
=
\loglik_\infty(r_*)
\approx-3.245.
\]

Numerical maximization of the Cauchit log-likelihood gives
\[
\hat\beta_{\mathrm{Cauchit}}\approx(2.838,-1.055),
\qquad
\loglik(\hat\beta_{\mathrm{Cauchit}})\approx -2.783.
\]
Since \(-2.783>-3.245\), this finite value strictly dominates all regular-variation boundary limits. Consequently the Cauchit model has a finite maximum, even though the same data are completely pairwise separated and cause divergence for the logit model. This contrasts with standard separation criteria based on geometric arguments, since the finite MLE lies outside the recession cone (note $\hat{\beta}_2 < 0$, whereas the cone requires $v_2 \ge 0$).

The geometric justification for this finite solution lies in the six middle observations. The competitor's covariate vector is $z_{i2} = (1, 3)$ $(i=2,\ldots,7)$. While the recession cone requires positive parameters to satisfy the global separation inequalities, a negative $\beta_2$ locally benefits the fit by suppressing the competitor's score ($\beta_1 + 3\beta_2$), thereby increasing the probability of the observed outcome. In the regular variation regime, the penalty for violating the separation inequalities grows slowly enough that this local improvement creates a global peak at finite values.

This illustrates the finite-MLE side of nondecisive separation. Data satisfying the escape condition, but not quasi-one-vs-all separation, may possess a finite MLE for regularly varying models, even when the same data cause complete divergence for logistic regression.

The same behaviour occurs in real data. For the wine bitterness data of \citet{Randall}, with the five ratings collapsed to three categories, the proportional adjacent-category logit model has a recession direction, whereas the corresponding Cauchit model attains $\ell(\hat\beta)=-30.815$, above its full boundary supremum $B^*=-31.420$. Details are given in Section~S2.2 of the Supplementary Material.

\section{Discussion}

Reference-ratio categorical regression shows that separation is not purely a geometric phenomenon. The same separating cone can have different likelihood consequences depending on the tail variation of the odds transform. Rapidly varying odds recover the classical behaviour in which strict advantages overwhelm competitors, while regularly varying odds retain positive same-tail boundary contributions and can therefore produce finite plateaux.

This distinction has direct software implications: numerical convergence in fitting routines such as \pkg{GLMcat} \citep{glmcat} should not be treated as evidence that a finite MLE exists, while geometric separation checks in \pkg{detectseparation} \citep{detectseparation} and \pkg{divoRce} \citep{divorce,sep-software} require a tail-aware interpretation for nonlogistic reference-ratio models.

This gives a simple interpretation of heavy-tailed categorical links. They do not remove separation geometry. Instead, they weaken its effect far out in the parameter space. The resulting plateau regime is nondecisive: geometry identifies possible escape directions, but it does not determine whether the boundary or an interior likelihood value wins. Under regular variation, Proposition~\ref{prop:rr-ray-completion} shows that the full boundary supremum is obtained from affine rays, so curved or multiscale paths cannot raise it. The examples show both failures of the classical dichotomy: completely separated data can possess a finite maximizer, whereas data with pairwise overlap can have an unattained boundary supremum.

The concavity result shows the complementary limitation. Global concavity in the full normalized-coordinate reference-ratio family forces affine log odds, and hence the multinomial logit. Thus the robustness gained from regularly varying links comes at the price of leaving the globally concave exponential-family class. This tradeoff is structural and should be considered when choosing between computational convenience and separation robustness.

\section*{Replication material}

The \texttt{R} code used to reproduce Fig.~\ref{fig:plt_sim}, Fig.~S1 and the reported numerical optimizations \ReplicationLocation. The Supplementary Material contains the proofs of all lemmas and of Propositions~\ref{prop:rr-ray-completion} and~\ref{prop:fixedref}, a two-observation example showing that quasi-complete separation is decisive for logit but not for probit or complementary log-log, and a real-data intermediate-plateau example based on the wine bitterness experiment.

\GenerativeAIDeclaration

\bibliographystyle{plainnat}
\bibliography{poly_ref}

\clearpage
\appendix
\setcounter{theorem}{0}
\setcounter{lemma}{0}
\setcounter{proposition}{0}
\setcounter{corollary}{0}
\setcounter{definition}{0}
\setcounter{equation}{0}
\setcounter{figure}{0}
\setcounter{table}{0}
\renewcommand{\thetheorem}{S\arabic{theorem}}
\renewcommand{\thelemma}{S\arabic{lemma}}
\renewcommand{\theproposition}{S\arabic{proposition}}
\renewcommand{\thecorollary}{S\arabic{corollary}}
\renewcommand{\thedefinition}{S\arabic{definition}}
\renewcommand{\theequation}{S\arabic{equation}}
\renewcommand{\thefigure}{S\arabic{figure}}
\renewcommand{\thetable}{S\arabic{table}}
\renewcommand{\thesection}{S\arabic{section}}
\begin{center}
{\Large\bfseries Supplementary Material}
\end{center}
\section{Proofs}
\label{sec:supp-proofs}

\subsection{Proof of Lemma~1}

\begin{restated}{Lemma~1 (Step $\Rightarrow$ scaling)}
If $\tau$ satisfies \textup{(2)} and is ultimately increasing, then $\tau$ is rapidly varying in the sense of \textup{(1)}.
\end{restated}

\begin{proof}
Fix $a>b\ge 0$ and take $\Delta,c,S$ from \textup{(2)}. If $b=0$, the numerator $\tau(0)$ is fixed and the denominator $\tau(at)$ diverges by iterating \textup{(2)}, so the ratio tends to zero. We therefore assume $b>0$. For large $t$,
write $at-bt=N_t\Delta+r_t$ with $N_t=\big\lfloor \tfrac{(a-b)t}{\Delta}\big\rfloor\to\infty$ and $r_t\in[0,\Delta)$. For $t$ large enough so that $bt\ge S$,
\[
\frac{\tau(bt)}{\tau(at)}
=\frac{\tau(bt)}{\tau(bt+\Delta)}\frac{\tau(bt+\Delta)}{\tau(bt+2\Delta)}
\cdots
\frac{\tau(bt+N_t\Delta)}{\tau(bt+N_t\Delta+r_t)}.
\]
By \textup{(2)}, each full-step factor is $\le c^{-1}$, hence their product is $\le c^{-N_t}$. Since $\tau$ is ultimately increasing and $r_t\ge 0$, the last (partial-step) factor is $\le 1$. Therefore
\[
\frac{\tau(bt)}{\tau(at)} \,\le\, c^{-N_t}\xrightarrow[t\to\infty]{}0,
\]
which is \textup{(1)}.
\end{proof}

\subsection{Proof of Lemma~2}

\begin{restated}{Lemma~2 (Asymptotic ratios for rapidly varying models)}
For a rapidly varying model, let $a, b \in \mathbb{R}$ satisfy $a > b$, and let $c,d\in \R$. Then
\[
\lim_{t\to\infty} \frac{\tau(b t + d)}{\tau(a t + c)} = 0.
\]
\end{restated}

\begin{proof}
Let $R(t)=\tau(bt+d)/\tau(at+c)$. If $a>b>0$, choose $M$ with $|c|,|d|\le M$. For all large $t$,
\[
\frac{\tau(bt-M)}{\tau(at+M)}
\le R(t)\le
\frac{\tau(bt+M)}{\tau(at-M)}
=\frac{\tau(x_t)}{\tau(q_t x_t)},
\]
where $x_t=bt+M$. Also \(q_t=(at-M)/(bt+M)\), so \(q_t\to a/b>1\). Choose
\(q_0\in(1,a/b)\). Then \(q_t\ge q_0\) for all sufficiently large \(t\).
By ultimate monotonicity,
\[
\frac{\tau(x_t)}{\tau(q_t x_t)} \le \frac{\tau(x_t)}{\tau(q_0 x_t)}.
\]
The right-hand side tends to zero by rapid variation, applied with the fixed
scale factor \(q_0>1\). Since \(R(t)\ge0\), the upper bound tends to zero and
therefore \(R(t)\to0\).

If $a>0\ge b$, the denominator diverges and the numerator is bounded or tends to zero. If $a=0>b$, the denominator is finite and positive, while the numerator tends to zero. Finally, if $0>a>b$, write $\tilde\tau(u)=1/\tau(-u)$. Then
\[
R(t)=\frac{\tilde{\tau}(|a|t-c)}{\tilde{\tau}(|b|t-d)},
\]
and the right-hand side tends to zero because $\tilde\tau$ is rapidly varying and $0<|a|<|b|$. These cases exhaust $a>b$.
\end{proof}

\subsection{Proof of Lemma~3}

\begin{restated}{Lemma~3 (Asymptotic ratios for regularly varying models)}
For a regularly varying model with right-tail index ${\rho_r} > 0$ and left-tail index ${\rho_l} > 0$, let $a, b \in \mathbb{R}$ satisfy $a > b$, and let $c,d\in\R$. The limit of the ratio $R(t) = \frac{\tau(b t + d)}{\tau(a t + c)}$ as $t \to \infty$ is as follows.
\begin{enumerate}
    \item If $a > b > 0$, then $\lim_{t\to\infty} R(t) = \left(\frac{b}{a}\right)^{\rho_r} \in (0, 1)$.
    \item If $0 > a > b$, then $\lim_{t\to\infty} R(t) = \left(\frac{|a|}{|b|}\right)^{\rho_l} \in (0, 1)$.
    \item If $a > 0 \ge b$ or $a = 0, b < 0$, then $\lim_{t\to\infty} R(t) = 0$.
\end{enumerate}
\end{restated}

\begin{proof}
Let $R(t) = \tau(b t + d) / \tau(a t + c)$. Since $d, c$ are bounded, in the cases where a and b share the same sign (both positive or both negative), the asymptotic behavior is determined by the linear terms $R(t) \sim \tau(bt)/\tau(at)$ as $t \to \infty$. This follows from the uniform convergence theorem for slowly varying functions \cite[Thm~1.2.1]{bingham1989regular}. For \(a,b<0\), the same argument is applied to the left-tail transform \(\tilde\tau(s)=1/\tau(-s)\).

\begin{enumerate}
    \item If $a > b > 0$, both arguments tend to $+\infty$. Using the property of regularly varying functions in Equation~\textup{(3)},
    \[
    \lim_{t\to\infty} R(t) = \lim_{t\to\infty} \frac{\tau(bt)}{\tau(at)} = \left(\frac{b}{a}\right)^{\rho_r}.
    \]
    Since $a>b>0$, this limit is a positive constant strictly less than 1.

    \item If $0 > a > b$, both arguments tend to $-\infty$. We use the left-tail property $\tau(s) = 1/\tilde{\tau}(-s)$.
    \[
    R(t) \sim \frac{1/\tilde{\tau}(-bt)}{1/\tilde{\tau}(-at)} = \frac{\tilde{\tau}(-at)}{\tilde{\tau}(-bt)} = \frac{\tilde{\tau}(|a|t)}{\tilde{\tau}(|b|t)}.
    \]
    Let $a' = |a|$ and $b' = |b|$. The condition $0 > a > b$ implies $0 < a' < b'$. Both arguments of $\tilde{\tau}$ tend to $+\infty$. By the same logic as in part 1, applied to $\tilde{\tau}$ with index $\rho_l$,
    \[
    \lim_{t\to\infty} R(t) = \left(\frac{a'}{b'}\right)^{\rho_l} = \left(\frac{|a|}{|b|}\right)^{\rho_l}.
    \]
    Since $|a| < |b|$, this limit is a positive constant strictly less than 1.
    
    \item If $a > 0 \ge b$ or $a = 0, b < 0$:
    \begin{enumerate}
        \item[(i)] If $a > 0 \ge b$, the denominator $\tau(at+c) \to \infty$. The numerator $\tau(bt+d)$ either tends to a finite constant (if $b=0$) or to $0$ (if $b<0$). In both subcases, the limit of the ratio is $0$.
        \item[(ii)] If $a = 0$ and $b < 0$, the denominator $\tau(0\cdot t + c) \to \tau(c)$, a finite positive constant, whereas the numerator $\tau(bt+d) \to 0$ as $bt \to -\infty$. The ratio therefore tends to $0$.
    \end{enumerate}
    In all cases, the limit is 0.
\end{enumerate}
\end{proof}

\subsection{Proof of Lemma~4}

\begin{restated}{Lemma~4 (Ties for regularly varying models)}
Let $\tau$ be the odds transform of a regularly varying model with indices ${\rho_r}, \rho_l>0$.
Fix $A\neq 0$ and bounded $c,d\in\R$. Then
\[
\frac{\tau(A t + d)}{\tau(A t + c)}
\,\xrightarrow[t\to\infty]{}\, 1.
\]
\end{restated}

\begin{proof}
If $A>0$, then $A t + d \to +\infty$ and $A t + c \to +\infty$. Regular variation and the uniform convergence theorem for slowly varying functions give
\[
\frac{\tau(A t + d)}{\tau(A t + c)}
= \Big(\frac{A t + d}{A t + c}\Big)^{\rho_r}
  \frac{L(A t + d)}{L(A t + c)}
\,\xrightarrow[t\to\infty]{}\, 1^{\rho_r}\cdot 1 = 1,
\]
because $d$ and $c$ are fixed and $L$ is slowly varying.
For $A<0$, the proof follows analogously using $A=-|A|$ and left-tail odds
\(
\tilde\tau(u) := 1/\tau(-u).
\)
\end{proof}

\subsection{Proof of Proposition~4}

\begin{restated}{Proposition~4 (Affine-ray completion of the boundary)}
For a regularly varying fixed-reference model on the finite-dimensional
estimable space \(\calX\), every subsequential limit of \(L(\beta_n)\) along
a sequence satisfying \(\|\beta_n\|\to\infty\) belongs to the closure of
\[
\left\{
L_\infty(\beta_0,v):
\beta_0\in\calX,\;
v\in\calX\setminus\{0\}
\right\}.
\]
Consequently, when $\dim\mathcal X\ge1$,
\[
 L_{\partial}
 =\lim_{R\to\infty}
      \sup_{\substack{\beta\in\mathcal X\\\|\beta\|\ge R}}L(\beta)
 =\max_{\substack{v\in\mathcal X\\\|v\|=1}}\;
      \sup_{\beta_0\in\mathcal X}L_\infty(\beta_0,v).
\]
The same identity holds for the log-likelihood, with
\(\log 0=-\infty\).
\end{restated}

\begin{proof}
Let $\beta_n\in\calX$ with $t_n=\|\beta_n\|\to\infty$. Passing to a subsequence, we may assume that $L(\beta_n)$ converges and, since the unit sphere of $\calX$ is compact, that $\beta_n/t_n\to v$ with $\|v\|=1$. Writing $a_{ik}=z_{ik}^\T v$, this gives
\[
z_{ik}^\T\beta_n=t_n\{a_{ik}+o(1)\}\qquad(n\to\infty).
\]
We compare $L(\beta_n)$ with $L_\infty(\beta_n,v)$, the likelihood limit along the ray $s\mapsto\beta_n+sv$ with $n$ held fixed. Both are products of finitely many probabilities in $[0,1]$, and $|\prod_i x_i-\prod_i y_i|\le\sum_i|x_i-y_i|$ for such numbers. It therefore suffices to show that, for each row $i$, the two observed-category probabilities differ by $o(1)$. Fix $i$ and let $M_i=\max_k a_{ik}$. Since $z_{iJ}=0$, $M_i\ge0$.

If $M_i>0$, then $\tau(z_{ik}^\T\beta_n)/\tau(t_n)\to(a_{ik})_+^{\rho_r}$ for every $k$, where $(x)_+=\max(x,0)$. For $a_{ik}>0$, this follows by bracketing $z_{ik}^\T\beta_n$ between $(a_{ik}\pm\varepsilon)t_n$, using monotonicity of $\tau$ and right-tail regular variation, and letting $\varepsilon\downarrow0$. For $a_{ik}<0$, the numerator tends to zero while $\tau(t_n)\to\infty$. For $a_{ik}=0$, eventually $z_{ik}^\T\beta_n\le\varepsilon t_n$, so the limit superior is at most $\varepsilon^{\rho_r}$ for every $\varepsilon>0$. Dividing the numerator and denominator of $h_{i,y_i}(\beta_n)$ by $\tau(t_n)$ shows that it converges to $(a_{iy_i})_+^{\rho_r}/\sum_k(a_{ik})_+^{\rho_r}$, where the denominator is positive because $M_i>0$. By Lemmas~3 and~4, the limiting ratios along a ray depend only on the slopes. Hence this value is also the limit along the ray from any starting point in direction $v$, in particular from $\beta_n$.

If $M_i=0$, let $\mathcal Z_i=\{k:a_{ik}=0\}$, which contains the reference category, and let $\mathcal N_i=\{k:a_{ik}<0\}$. Along the sequence, $\delta_n=\sum_{k\in\mathcal N_i}\tau(z_{ik}^\T\beta_n)\to0$. Along the ray from $\beta_n$, the odds of the categories in $\mathcal N_i$ tend to zero, whereas the odds of the categories in $\mathcal Z_i$ remain equal to $\tau(z_{ik}^\T\beta_n)$. Let $D_n=\sum_{k\in\mathcal Z_i}\tau(z_{ik}^\T\beta_n)$, so that $D_n\ge\tau(0)=1$. If $y_i\in\mathcal Z_i$, the two probabilities are $u_n/(D_n+\delta_n)$ and $u_n/D_n$ with $u_n=\tau(z_{iy_i}^\T\beta_n)\le D_n$, and they differ by $u_n\delta_n/\{D_n(D_n+\delta_n)\}\le\delta_n$. If $y_i\in\mathcal N_i$, the sequence probability is at most $\delta_n$ and the ray probability is zero.

Hence $L(\beta_n)-L_\infty(\beta_n,v)\to0$, so $\lim_n L(\beta_n)$ is a limit of the affine-ray values $L_\infty(\beta_n,v)$. This proves the closure statement.

For the identity, every affine-ray limit is at most $L_\partial$, because $\|\beta_0+sv\|\to\infty$ as $s\to\infty$. Conversely, choose $\beta_n$ with $\|\beta_n\|\ge n$ and $L(\beta_n)\ge\sup_{\|\beta\|\ge n}L(\beta)-1/n$, so that $L(\beta_n)\to L_\partial$. Along a subsequence with limiting direction $v$, the comparison above gives
\[
L_\partial=\lim_{n\to\infty}L_\infty(\beta_n,v)\le\sup_{\beta_0\in\calX}L_\infty(\beta_0,v)\le L_\partial.
\]
Hence the maximum over unit directions is attained at $v$. Since $\log$ is increasing and continuous on $[0,1]$, with $\log0=-\infty$, the same identity holds for the log-likelihood.
\end{proof}

\subsection{Proof of Lemma~5}

\begin{restated}{Lemma~5 (Separation hierarchy)}
The geometric conditions are nested as follows.
\[
S1 \implies S2 \implies ESC.
\]
\end{restated}

\begin{proof}
To prove $S1 \implies S2$, let $v$ satisfy S1 and set $a_{ik}=z_{ik}^\T v$. For any observation $i$, S1 gives $a_{ij} \ge 0$ and $a_{ik} \le 0$ for all $k \ne j$, hence $a_{ij} \ge a_{ik}$ for all $k \ne j$. Moreover, since \(v\in\calX\setminus\{0\}\) and the likelihood is considered on the estimable span, not all category scores \(z_{ik}^\T v\) can vanish (otherwise \(v\) would be orthogonal to \(\calX\), and hence \(v=0\)). Under \(S1\), this implies that at least one inequality \(a_{ij}\ge a_{ik}\) is strict. Hence \(S2\) holds.

To prove $S2 \implies ESC$, let $v$ satisfy S2 and set $a_{ik}=z_{ik}^\T v$. For every observation $i$, S2 implies $a_{ij}=\max_k a_{ik}$. The escape condition then reduces to the exhaustive alternatives $a_{ij}>0$, $a_{ij}<0$ or $a_{ij}=0$.
\end{proof}

\subsection{Proof of Lemma~6}

\begin{restated}{Lemma~6}
Let $\loglik_i(\gamma_i)$ be the per-observation log-likelihood, and for fixed $i$ suppress the first index by writing $\gamma_r=\gamma_{ir}$ and $p_r=h_r(\gamma_i)=h_{ir}(\beta)$. The entries of the Hessian matrix $\nabla^2 \loglik_i(\gamma_i)$ are given by
\[
\frac{\partial^2 \loglik_i}{\partial \gamma_r \partial \gamma_s}
=\underbrace{\delta_{rs} \lodds''(\gamma_r) (\delta_{ry_i} - p_r)}_{\text{diagonal correction}}
\underbrace{- \lodds'(\gamma_r) \lodds'(\gamma_s) p_r (\delta_{rs} - p_s)}_{\text{variance term}}.
\]
Furthermore, for any vector $a \in \mathbb{R}^J$, the quadratic form of the variance term is
\[
\sum_{r=1}^J \sum_{s=1}^J a_r \left[ - \lodds'(\gamma_r) \lodds'(\gamma_s) p_r (\delta_{rs} - p_s) \right] a_s = - \mathrm{Var}_{p_i}(R),
\]
where $R$ is a discrete random variable taking value $w_k = \lodds'(\gamma_k) a_k$ with probability $p_k$.
\end{restated}

\begin{proof}
We find the Hessian by differentiating the gradient, $G_r = \frac{\partial \loglik_i}{\partial \gamma_r} = \lodds'(\gamma_r) ( \delta_{ry_i} - p_r )$, using the product rule $H_{rs} = \frac{\partial G_r}{\partial \gamma_s} = \frac{\partial (\lodds'(\gamma_r))}{\partial \gamma_s} ( \delta_{ry_i} - p_r ) + \lodds'(\gamma_r) \frac{\partial ( \delta_{ry_i} - p_r )}{\partial \gamma_s}$.

The first part of the product rule gives the diagonal correction
\[
\frac{\partial (\lodds'(\gamma_r))}{\partial \gamma_s} ( \delta_{ry_i} - p_r ) = \delta_{rs} \lodds''(\gamma_r) (\delta_{ry_i} - p_r).
\]
The second part gives the variance term. We need $\frac{\partial (-p_r)}{\partial \gamma_s}$. Using the chain rule with $u_s = \lodds(\gamma_s)$ and the derivative of the softmax function, we know $\frac{\partial p_r}{\partial \gamma_s} = \frac{\partial p_r}{\partial u_s} \frac{d u_s}{d \gamma_s} = \left( p_r(\delta_{rs} - p_s) \right) \lodds'(\gamma_s)$.
Thus, the second part of the product rule is
\[
\lodds'(\gamma_r) \frac{\partial (-p_r)}{\partial \gamma_s} = - \lodds'(\gamma_r) \lodds'(\gamma_s) p_r(\delta_{rs} - p_s).
\]
Summing these two parts gives the Hessian formula.

To prove the quadratic form, let $H^V$ be the variance term matrix. Then
\begin{align*}
a^\T H^V a &= \sum_r \sum_s a_r \left[ - \lodds'(\gamma_r) \lodds'(\gamma_s) p_r (\delta_{rs} - p_s) \right] a_s \\
&= - \sum_r \sum_s \left( \lodds'(\gamma_r) a_r \right) \left( \lodds'(\gamma_s) a_s \right) p_r (\delta_{rs} - p_s) \\
&= - \sum_r \sum_s w_r w_s p_r (\delta_{rs} - p_s) \quad (\text{letting } w_k = \lodds'(\gamma_k) a_k) \\
&= - \left[ \sum_r \sum_s w_r w_s p_r \delta_{rs} - \sum_r \sum_s w_r w_s p_r p_s \right] \\
&= - \left[ \sum_r w_r^2 p_r - \left(\sum_r w_r p_r\right) \left(\sum_s w_s p_s\right) \right] \\
&= - \left[ \E[R^2] - (\E[R])^2 \right] = - \mathrm{Var}_{p_i}(R).
\end{align*}
Since $\mathrm{Var}_{p_i}(R) \ge 0$, the variance term is always negative semidefinite.
\end{proof}

\subsection{Proof of Proposition~6}

\begin{restated}{Proposition~6}
Let \(m=J-1\ge2\), and let \(\tau\) be twice continuously differentiable with \(\tau'(s)>0\) for every \(s\in\R\) and \(\tau(s)\to\infty\) as \(s\to\infty\). Then the fixed-reference log-likelihood is globally concave in the regression parameter for every design and every response vector if and only if
\[
-\frac{\{\lodds'(s)\}^{2}}{1+m\tau(s)}
\leq
\lodds''(s)
\leq 0
\qquad (s\in\R).
\]
\end{restated}

Write \(\lodds=\log\tau\). For
\(\eta=(\eta_1,\ldots,\eta_m)^\T\), define
\[
h_j(\eta)
=
\frac{\tau(\eta_j)}
     {1+\sum_{k=1}^m\tau(\eta_k)},
\qquad j=1,\ldots,m,
\]
and
\[
h_J(\eta)
=
\frac{1}
     {1+\sum_{k=1}^m\tau(\eta_k)}.
\]

We show that the following statements are equivalent; the equivalence of
(b) and (c) is Proposition~6.
\begin{enumerate}
\item[(a)]
For every \(y\in[J]\), the function
\(\eta\mapsto\log h_y(\eta)\) is globally concave on
\(\mathbb{R}^m\).

\item[(b)]
For every design and every response vector, the corresponding
fixed-reference log-likelihood is globally concave in the regression
parameter.

\item[(c)]
Inequality \textup{(4)} holds for every
\(s\in\mathbb{R}\).
\end{enumerate}

\begin{proof}
The equivalence of (a) and (b) follows because affine composition and
summation preserve concavity, while the reverse implication is obtained by
taking a single observation with the moving-score vector itself as parameter.
It therefore suffices to prove the equivalence of (a) and (c).

Write
\[
D(\eta)=1+\sum_{k=1}^m\tau(\eta_k),
\qquad
p_k(\eta)=\frac{\tau(\eta_k)}{D(\eta)}.
\]
Restricting the Hessian formula in Lemma~6 to \(\gamma_J=0\), with \(y=J\), gives
\begin{equation}
H(\eta) :=
-\nabla^2\log h_J(\eta)
=
A(\eta)-u(\eta)u(\eta)^\T,
\label{eq:fixed-reference-ref-hessian}
\end{equation}
where
\[
A(\eta)
=
\operatorname{diag}
\left\{
p_k(\eta)
\left[
\{\lodds'(\eta_k)\}^2+\lodds''(\eta_k)
\right]
\right\}_{k=1}^m
\]
and
\[
u_k(\eta)=p_k(\eta)\lodds'(\eta_k).
\]
Moreover,
\[
\log h_j(\eta)=\log h_J(\eta)+\lodds(\eta_j),
\qquad j=1,\ldots,m,
\]
so
\begin{equation}
-\nabla^2\log h_j(\eta)
=
A(\eta)-u(\eta)u(\eta)^\T
-\lodds''(\eta_j)e_je_j^\T.
\label{eq:fixed-reference-moving-hessian}
\end{equation}
For necessity, the argument used in the proof of
Theorem~2 to obtain \(\lodds''(s)\leq0\) applies without
change. Fix a moving score at \(s\) and send a different moving score to
\(+\infty\). This is possible because \(m\geq2\).

To obtain the remaining necessary inequality, set
\(\eta_1=\cdots=\eta_m=s\). Concavity of the reference-category
log-probability implies
\[
0
\leq
\mathbf 1^\T
\{-\nabla^2\log h_J(s\mathbf 1)\}
\mathbf 1.
\]
A direct simplification gives
\[
0
\leq
\frac{m\tau(s)}{1+m\tau(s)}
\left[
\lodds''(s)
+
\frac{\{\lodds'(s)\}^2}{1+m\tau(s)}
\right],
\]
and therefore
\[
\lodds''(s)
\geq
-\frac{\{\lodds'(s)\}^2}{1+m\tau(s)}.
\]
This proves the necessity of condition (c).

We now prove sufficiency. Assume condition (c). For every \(k\),
\begin{align*}
d_k(\eta)
&=
\{\lodds'(\eta_k)\}^2+\lodds''(\eta_k)\\
&\geq
\{\lodds'(\eta_k)\}^2
\frac{m\tau(\eta_k)}{1+m\tau(\eta_k)}
>0.
\end{align*}
Thus \(A(\eta)\) is positive definite. For a positive definite matrix
\(A\), the rank-one criterion gives
\[
A-uu^\T\succeq0
\quad\Longleftrightarrow\quad
u^\T A^{-1}u\leq1.
\]
In the present case,
\begin{align}
u(\eta)^\T A(\eta)^{-1}u(\eta)
&=
\sum_{k=1}^m
\frac{p_k(\eta)\{\lodds'(\eta_k)\}^2}
     {d_k(\eta)}
\nonumber\\
&=
\frac{1}{D(\eta)}
\sum_{k=1}^m
\frac{\tau(\eta_k)\{\lodds'(\eta_k)\}^2}
     {d_k(\eta)}
\nonumber\\
&=
\frac{1}{D(\eta)}
\left[
\sum_{k=1}^m\tau(\eta_k)
+
\sum_{k=1}^m
\frac{-\tau(\eta_k)\lodds''(\eta_k)}
     {d_k(\eta)}
\right].
\label{eq:fixed-reference-rank-one}
\end{align}
The lower inequality in condition (c) is equivalent to
\[
\frac{-\tau(\eta_k)\lodds''(\eta_k)}
     {d_k(\eta)}
\leq
\frac{1}{m}.
\]

Therefore,
\[
u(\eta)^\T A(\eta)^{-1}u(\eta)
\leq
\frac{\sum_{k=1}^m\tau(\eta_k)+1}{D(\eta)}
=1.
\]
Hence
\[
H(\eta)=A(\eta)-u(\eta)u(\eta)^\T\succeq0,
\]
and \(\log h_J\) is concave. Finally, the upper inequality in (c) gives
\(\lodds''(\eta_j)\leq0\), so
\[
H(\eta)-\lodds''(\eta_j)e_je_j^\T
=
H(\eta)+\{-\lodds''(\eta_j)\}e_je_j^\T
\succeq0.
\]
By \eqref{eq:fixed-reference-moving-hessian}, every moving-category
log-probability is also concave. This proves (a).
\end{proof}

The criterion has two consequences that connect fixed-reference concavity to
the tail classification in the main paper. First, it continues to rule out
regularly varying odds. Indeed, the upper inequality in
\textup{(4)} implies that \(\lodds'\) is
nonincreasing. Under the standing assumption \(\lodds'(s)>0\), there is
therefore a limit
\[
\lodds'(s)\longrightarrow a_+\geq0,
\qquad s\to\infty.
\]
The lower inequality can be written as
\[
\frac{\lodds''(s)}
     {\{\lodds'(s)\}^2}
\geq
-\frac{1}{1+m\exp\{\lodds(s)\}}.
\]
Since \(\lodds'(s)>0\), multiplying by \(\lodds'(s)\) gives
\[
\frac{d}{ds}\log\lodds'(s)
=
\frac{\lodds''(s)}{\lodds'(s)}
\geq
-\frac{\lodds'(s)}
       {1+m\exp\{\lodds(s)\}}.
\]
Therefore, for any fixed \(s_0\),
\[
\log\lodds'(s)-\log\lodds'(s_0)
\geq
-\int_{s_0}^{s}
\frac{\lodds'(t)}
     {1+m\exp\{\lodds(t)\}}\,dt.
\]
With the substitution \(u=\lodds(t)\), this becomes
\[
\log\lodds'(s)-\log\lodds'(s_0)
\geq
-\int_{\lodds(s_0)}^{\lodds(s)}
\frac{du}{1+m e^u}.
\]
The integral remains bounded as \(s\to\infty\), since
\[
\int_{\lodds(s_0)}^\infty
\frac{du}{1+m e^u}
<\infty.
\]
Hence \(\lodds'(s)\) is bounded away from zero in the right tail. Since the
upper inequality in \textup{(4)} gives
\(\lodds''(s)\leq0\), the function \(\lodds'\) is nonincreasing and therefore
\[
\lodds'(s)\longrightarrow a_+
\]
for some \(a_+>0\). It follows that
\[
\lodds(s)=a_+s+o(s),
\qquad s\to\infty,
\]
or equivalently,
\[
\tau(s)=\exp\{a_+s+o(s)\}.
\]
Thus, for \(J\ge3\), a globally concave fixed-reference link has asymptotically
linear log odds in the right tail. In particular, a regularly varying odds transform
cannot satisfy the fixed-reference global concavity criterion, since regular
variation implies
\[
\frac{\log\tau(s)}{s}\longrightarrow0.
\]

Second, the finite-dimensional exceptions disappear if concavity is required
uniformly over the number of categories. Suppose that the same odds transform
\(\tau\) gives a globally concave fixed-reference likelihood for every
\(m\geq2\). Then, for each fixed \(s\),
\[
-\frac{\{\lodds'(s)\}^2}{1+m\tau(s)}
\leq
\lodds''(s)
\leq0
\qquad\text{for every }m\geq2.
\]
Letting \(m\to\infty\) yields \(\lodds''(s)\geq0\), and hence
\[
\lodds''(s)=0
\qquad\text{for every }s\in\mathbb{R}.
\]
Therefore \(\lodds\) is affine and
\[
\tau(s)=Ce^{as}
\]
for some \(C>0\) and \(a>0\). Multinomial logit is thus the unique link whose
fixed-reference global concavity is preserved uniformly over the number of
categories.

\section{Additional examples}
\label{sec:supp-examples}

\subsection{Quasi-complete separation under rapid tails}

\label{sec:supp-rapid-quasi-example}

Quasi-complete pairwise separation is decisive for logit, but not for probit
or complementary log-log. Consider a fixed-reference model with
\(n=2\), \(J=3\), and \(\beta=(x,w)^\T\). Let
\[
(y_1;z_{11},z_{12},z_{13})
=
\bigl(1;(2,-1),(2,0),0\bigr),
\]
and
\[
(y_2;z_{21},z_{22},z_{23})
=
\bigl(2;(2,1),(2,3),0\bigr).
\]
The structure vectors are
\[
(0,-1),\qquad (2,-1),\qquad (0,2),\qquad (2,3).
\]
Their inner products with \(v=(1,0)^\T\) are \(0,2,0,2\), so
quasi-complete separation holds. Complete separation is impossible because
the first and third structure inequalities would require respectively
\(v_2<0\) and \(v_2>0\). Quasi-one-vs-all separation is also impossible:
its first-row inequalities imply \(v_1\leq0\) and \(v_2\leq2v_1\), whereas
the observed-category inequality in the second row requires
\(2v_1+3v_2\geq0\); together these force \(v=0\).

For an odds transform \(\tau\), the likelihood is
\[
L(x,w)
=
\frac{\tau(2x-w)}
     {1+\tau(2x-w)+\tau(2x)}
\,
\frac{\tau(2x+3w)}
     {1+\tau(2x+w)+\tau(2x+3w)}.
\]
We now determine its full boundary envelope. The quasi-separating cone is
exactly \(\{t(1,0)^\T:t>0\}\). Hence, along any divergent sequence with a
nonzero likelihood limit, necessarily
\[
x\to\infty,
\qquad
w=o(x).
\]
Every other limiting direction gives a strict competitor advantage in at
least one row and therefore likelihood limit zero by rapid variation.

For probit odds
\[
\tau_{\rm P}(s)=\frac{\Phi(s)}{\Phi(-s)},
\qquad
\lambda_{\rm P}(s)=\log\tau_{\rm P}(s)
=
\frac{s^2}{2}+\log s+\frac12\log(2\pi)+o(1).
\]
After passing to a subsequence, either \(xw\to\pm\infty\), in which case one
row probability tends to zero, or \(xw\to c\in\mathbb{R}\). In the latter
case,
\[
\frac{\tau_{\rm P}(2x)}
     {\tau_{\rm P}(2x-w)}
\longrightarrow e^{2c},
\qquad
\frac{\tau_{\rm P}(2x+w)}
     {\tau_{\rm P}(2x+3w)}
\longrightarrow e^{-4c}.
\]

For the centered complementary log-log odds
\[
\tau_{\rm C}(s)=\exp\{(\log 2)e^s\}-1,
\qquad
\lambda_{\rm C}(s)=(\log 2)e^s+o(1).
\]
The corresponding critical scale is \(e^{2x}w\). If
\(e^{2x}w\to d\in\mathbb{R}\), then
\[
\frac{\tau_{\rm C}(2x)}
     {\tau_{\rm C}(2x-w)}
\longrightarrow e^{(\log 2)d},
\qquad
\frac{\tau_{\rm C}(2x+w)}
     {\tau_{\rm C}(2x+3w)}
\longrightarrow e^{-2(\log 2)d}.
\]
If \(e^{2x}w\to\pm\infty\), one row probability again tends to zero.

Thus both links have the same family of nonzero boundary limits,
\[
B(q)
=
\frac{q^2}{(1+q)(1+q^2)},
\qquad q>0.
\]
Since
\[
\frac{d}{dq}\log B(q)
=
-\frac{q^3-q-2}{q(1+q)(1+q^2)},
\]
the unique maximum occurs at the positive root
\(q_*\approx1.5214\) of \(q^3-q-2=0\), and
\[
B^*
=
B(q_*)
\approx
0.277.
\]
Finite points exceed this full boundary supremum with
\[
L_{\rm P}(1,1/5)\approx 0.284>B^*,
\qquad
L_{\rm C}(4/5,3/20)\approx 0.296 >B^*.
\]
Since every divergent sequence has likelihood limsup at most \(B^*\),
continuity implies that both likelihoods attain their global maxima at finite
parameter values. On the same data, the logit likelihood has no finite
maximizer by the usual quasi-complete-separation argument. Hence
quasi-complete separation is genuinely nondecisive for probit and
complementary log-log reference-ratio models.

\subsection{Wine bitterness data}

\label{sup:wine-real-example}

Section~5 uses a deliberately small synthetic example to isolate the
mechanism by which regularly varying reference-ratio links can retain a
finite maximum under separation. We now show the same mechanism in a
published sensory experiment. The wine bitterness data of \citet{Randall}
are a standard ordinal-regression benchmark and are distributed with the
\pkg{ordinal} package \citep{ordinal}. The experiment has two controlled treatment factors,
temperature (\texttt{cold}, \texttt{warm}) and skin contact
(\texttt{no}, \texttt{yes}), and \(72\) ordinal bitterness ratings. To focus
on the two tails of the sensory scale, we collapse the original five
ratings to
\[
  Y=1,\qquad Y=2\text{-}4,\qquad Y=5.
\]
This gives the table
\[
\begin{array}{llccc}
\toprule
\text{temperature} & \text{contact} & Y=1 & Y=2\text{-}4 & Y=5\\
\midrule
\text{cold} & \text{no}  & 4 & 14 & 0\\
\text{cold} & \text{yes} & 1 & 17 & 0\\
\text{warm} & \text{no}  & 0 & 16 & 2\\
\text{warm} & \text{yes} & 0 & 13 & 5\\
\bottomrule
\end{array}
\]
The absence of low ratings under warm temperature and high ratings under
cold temperature is the familiar sparse-cell pattern that creates
separation in categorical regression.

We use category \(3\), corresponding to the collapsed rating \(Y=5\), as
the reference category. Let \(x_{i1}=1\) indicate warm temperature and
\(x_{i2}=1\) indicate skin contact. In parameter order
\[
  \beta=(\alpha_1,\alpha_2,\theta_{\rm temp},\theta_{\rm contact})^\T,
\]
we use the parallel adjacent-category score representation
\[
  z_{i1}=(1,1,-2x_{i1},-2x_{i2})^\T,\qquad
  z_{i2}=(0,1,-x_{i1},-x_{i2})^\T,\qquad
  z_{i3}=0 .
\]
With \(\tau(s)=\exp(s)\), this is the proportional adjacent-category logit
model written in normalized reference-ratio form. Replacing
\(\tau(s)=\exp(s)\) by the centered Cauchit odds gives the corresponding
regularly varying reference-ratio model with the same scores.

The pairwise-separation cone is one-dimensional. The inequalities
\((z_{i,y_i}-z_{ik})^\T v\geq0\), over all observed rows and competitors
\(k\), imply
\[
  v_1=0,\qquad v_4=0,\qquad v_2=v_3\geq0,
\]
and hence
\[
  C=\{t(0,1,1,0)^\T:t\geq0\}.
\]
Thus the logit model has a recession direction. In particular, along
\(\beta(t)=\beta_0+t(0,1,1,0)^\T\), the logit likelihood increases towards
an unattained limiting value for every finite starting point \(\beta_0\).

The nonzero escape cone for the Cauchit model is larger. Solving the escape
conditions for all observed rows gives, up to positive scaling,
\[
  v=(q,1,1,0)^\T,
  \qquad -1<q\leq1.
\]
The pairwise-separating ray is the special case \(q=0\). For \(-1<q<1\),
the cold-group slopes are \((q+1,1,0)\), whereas the warm-group slopes are
\((q-1,0,0)\). Writing
\[
p_0
=
\frac{\tau(\alpha_2-\theta_{\rm temp})}
     {1+\tau(\alpha_2-\theta_{\rm temp})},
\qquad
p_1
=
\frac{\tau(\alpha_2-\theta_{\rm temp}-\theta_{\rm contact})}
     {1+\tau(\alpha_2-\theta_{\rm temp}-\theta_{\rm contact})},
\]
the boundary log-likelihood is
\begin{align*}
B(q,p_0,p_1)
&=5\log\left(\frac{q+1}{q+2}\right)
 +31\log\left(\frac{1}{q+2}\right)\\
&\quad+16\log p_0+2\log(1-p_0)
 +13\log p_1+5\log(1-p_1).
\end{align*}
Maximizing over the finite offsets gives \(p_0=8/9\) and \(p_1=13/18\).
The resulting derivative with respect to \(q\) is
\[
B'(q)
=
\frac{5}{q+1}-\frac{36}{q+2}
=
\frac{-31q-26}{(q+1)(q+2)},
\]
so the unique maximizing direction is
\[
q_*=-\frac{26}{31},
\qquad
v_*\propto(-26,31,31,0)^\T.
\]
The full boundary maximum is therefore
\begin{align*}
B^* =5\log\frac{5}{36}+31\log\frac{31}{36}
 +16\log\frac89+2\log\frac19+13\log\frac{13}{18}+5\log\frac5{18}=-31.4200.
\end{align*}
The affine-ray completion result proved above shows that this optimization
also controls arbitrary curved and multiscale divergent sequences. At
\(q=-1\), an observed cold-category probability tends to zero. At \(q=1\),
the additional warm-category-\(1\) term is unobserved and can only decrease
the likelihood. Hence \(B^*\) is the full regular-variation boundary maximum.

Numerical maximization gives a finite Cauchit solution
\[
  \hat\beta_{\rm Cauchit}
  =
  (-24.453,\;32.380,\;28.712,\;2.864)^\T,
  \qquad
  \loglik(\hat\beta_{\rm Cauchit})=-30.8151 .
\]
This finite value exceeds the boundary maximum by approximately
\(0.6049\) log-likelihood units. It therefore dominates all boundary limits,
so the Cauchit likelihood attains its maximum at a finite point. A multi-start
optimization, including starts far along escape directions and large random
starts, returned the same finite solution.

\begin{figure}
\centering
\includegraphics[width=\textwidth]{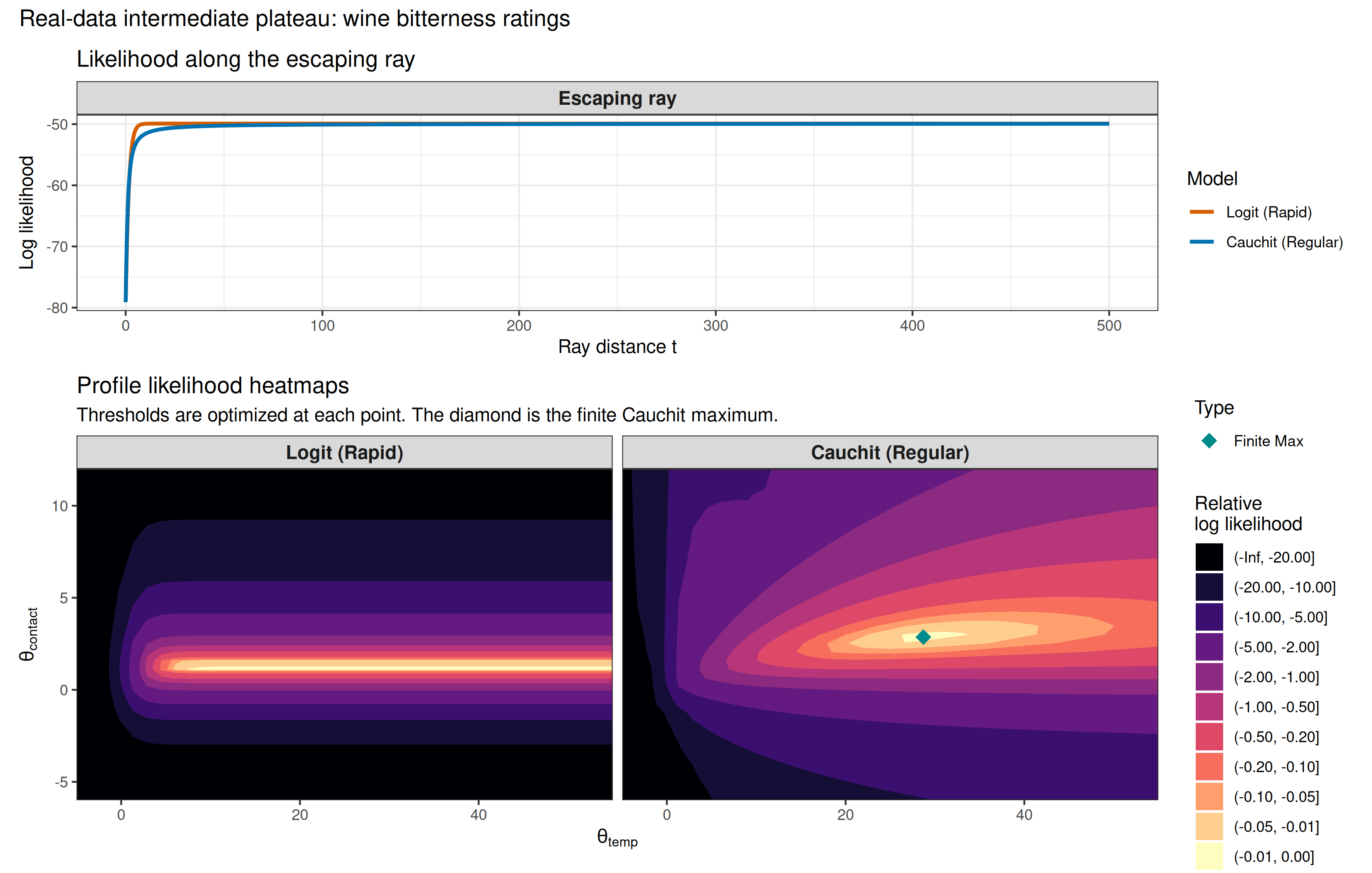}
\setlength{\figwidth}{\textwidth}
\caption{Real-data intermediate plateau for the collapsed wine bitterness
data. The upper panel shows log-likelihood profiles along the
pairwise-separating ray \(\beta(t)=t(0,1,1,0)^\T/\sqrt2\) started at the
origin. Both profiles approach \(72\log(1/2)\approx-49.91\), which is below
the boundary maximum \(B^*=-31.4200\)
because the starting point is not optimized. The lower panels show profiled log-likelihood heatmaps for logit and
Cauchit in the two slope coordinates. The diamond marks the finite Cauchit
maximum.}
\label{fig:wine-real}
\alttext{Log-likelihood profiles and heatmaps for the collapsed wine bitterness data. The profile plot compares logit and Cauchit along the pairwise-separating ray. The heatmaps show profiled log-likelihood values over the two treatment-effect coordinates. Logit improves toward the boundary, whereas Cauchit has a finite interior maximum marked by a diamond.}
\end{figure}

Figure~\ref{fig:wine-real} gives the same visual diagnostic as the
synthetic example in the main paper. The upper panel shows the likelihood
along the pairwise-separating ray \(q=0\), started at the origin, which is
also an escaping ray. Its plateau \(72\log(1/2)\approx-49.91\) lies below
\(B^*\) because the starting point, and hence the finite offsets, is not
optimized. The
lower panels show profiled likelihood heatmaps
in the two treatment-effect coordinates, with the two threshold parameters
optimized at each point. For logit, the highest values move towards the
boundary of the plotting window in the escaping direction. For Cauchit, the
maximum is attained at a finite interior point.

The calculations and figure are reproduced by \texttt{polyref.R}.

\end{document}